\documentclass[iop]{emulateapj}
\usepackage{natbib}
\usepackage{mathrsfs}
\usepackage{url}
\usepackage[normalem]{ulem}
\newcommand{\IUE}{{\it IUE}}
\newcommand{\HST}{{\it HST}}
\newcommand{\kms}{\ifmmode {\rm km~s}^{-1} \else km~s$^{-1}$\fi}
\newcommand{\ergs}{\ifmmode {\rm erg~ s}^{-1} \else erg~s$^{-1}$\fi}
\newcommand{\ergscm}{\ifmmode {\rm erg~s}^{-1} \else erg~s$^{-1}$ cm$^{-2}$\fi}
\newcommand{\Msun}{\ifmmode M_{\odot} \else $M_{\odot}$\fi }
\newcommand{\Lsun}{\ifmmode {\rm L}_{\odot} \else L$_{\odot}$\fi}
\newcommand{\qo}{\ifmmode q_{\rm o} \else $q_{\rm o}$\fi}
\newcommand{\Ho}{\ifmmode H_{\rm o} \else $H_{\rm o}$\fi}
\newcommand{\ho}{\ifmmode h_{\rm o} \else $h_{\rm o}$\fi}

\newcommand{\gtsim}{\raisebox{-.5ex}{$\;\stackrel{>}{\sim}\;$}}
\newcommand{\vFWHM}{\ifmmode v_{\mbox{\tiny FWHM}} \else
                    $v_{\mbox{\tiny FWHM}}$\fi}
\newcommand{\CCF}{\ifmmode F_{\it CCF} \else $F_{\it CCF}$\fi}
\newcommand{\ACF}{\ifmmode F_{\it ACF} \else $F_{\it ACF}$\fi}
\newcommand{\Halpha}{\ifmmode {\rm H}\alpha \else H$\alpha$\fi}
\newcommand{\Hbeta}{\ifmmode {\rm H}\beta \else H$\beta$\fi}
\newcommand{\Hgamma}{\ifmmode {\rm H}\gamma \else H$\gamma$\fi}
\newcommand{\Hdelta}{\ifmmode {\rm H}\delta \else H$\delta$\fi}
\newcommand{\Lya}{\ifmmode {\rm Ly}\alpha \else Ly$\alpha$\fi}
\newcommand{\Lyb}{\ifmmode {\rm Ly}\beta \else Ly$\beta$\fi}
\newcommand{\HeI}{\ifmmode {\rm He}\,{\sc i}\,\lambda5876 \else
	          He\,{\sc i}\,$\lambda5876$\fi}
\newcommand{\HeII}{\ifmmode {\rm He}\,{\sc ii}\,\lambda4686 \else 
	           He\,{\sc ii}\,$\lambda4686$\fi}
\newcommand{\hi}{\ifmmode \makebox{{\rm H\,}{\sc i}} \else H\,{\sc i}\fi}

\newcommand{\hei}{He\,{\sc i}}
\newcommand{\heii}{\ifmmode \makebox{{\rm He}\,{\sc ii}} \else He\,{\sc ii}\fi}

\newcommand{\feii}{Fe\,{\sc ii}}

\newcommand{\cii}{C\,{\sc ii}}
\newcommand{\ciii}{\ifmmode {\rm C}\,{\sc iii} \else C\,{\sc iii}\fi}
\newcommand{\ciiiex}{C\,{\sc iii}*}
\newcommand{\civ}{C\,{\sc iv}}
\newcommand{\Ni}{N\,{\sc i}}

\newcommand{\niiisf}{N\,{\sc iii}]}

\newcommand{\nivsf}{N\,{\sc iv}]}
\newcommand{\nv}{N\,{\sc v}}
\newcommand{\oi}{O\,{\sc i}}

\newcommand{\oiii}{O\,{\sc iii}}
\newcommand{\oiiisf}{O\,{\sc iii}]}

\newcommand{\oiv}{O\,{\sc iv}}

\newcommand{\Sizw}{Si\,{\sc ii}}
\newcommand{\siiv}{Si\,{\sc iv}}

\newcommand{\alii}{Al\,{\sc ii}}

\def\fake2{\hphantom{3}}

\shorttitle{AGN STORM. I. UV Observations}
\shortauthors{De Rosa et al.}

\begin{document}

\title{Space Telescope and Optical Reverberation Mapping Project.\\
I.\ Ultraviolet Observations of the Seyfert 1 Galaxy NGC\,5548 with\\ the Cosmic
Origins Spectrograph on {\it Hubble Space Telescope}}

\author{G.~De~Rosa\altaffilmark{1,2,3},
B.~M.~Peterson\altaffilmark{1,2},
J.~Ely\altaffilmark{3},
G.~A.~Kriss\altaffilmark{3,4},
D.~M.~Crenshaw\altaffilmark{5},
Keith~Horne\altaffilmark{6},
K.~T.~Korista\altaffilmark{7},
H.~Netzer\altaffilmark{8},
R.~W.~Pogge\altaffilmark{1,2},
P.~Ar\'{e}valo\altaffilmark{9},
A.~J.~Barth\altaffilmark{10},
M.~C.~Bentz\altaffilmark{5},
W.~N.~Brandt\altaffilmark{11},
A.~A.~Breeveld\altaffilmark{12}
B.~J.~Brewer\altaffilmark{13},
E.~Dalla~Bont\`{a}\altaffilmark{14,15},
A.~De~Lorenzo-C\'{a}ceres\altaffilmark{6},
K.~D.~Denney\altaffilmark{1,2,16},
M.~Dietrich\altaffilmark{17,18},
R.~Edelson\altaffilmark{19},
P.~A.~Evans\altaffilmark{20},
M.~M.~Fausnaugh\altaffilmark{1},
N.~Gehrels\altaffilmark{21},
J.~M.~Gelbord\altaffilmark{22,23},
M.~R.~Goad\altaffilmark{20},
C.~J.~Grier\altaffilmark{1,11},
D.~Grupe\altaffilmark{24},
P.~B.~Hall\altaffilmark{25},
J.~Kaastra\altaffilmark{26,27,28},
B.~C.~Kelly\altaffilmark{29},
J.~A.~Kennea\altaffilmark{11},
C.~S.~Kochanek\altaffilmark{1,2},
P.~Lira\altaffilmark{30},
S.~Mathur\altaffilmark{1,2},
I.~M.~M$^{\rm c}$Hardy\altaffilmark{31},
J.~A.~Nousek\altaffilmark{11},
A.~Pancoast\altaffilmark{29},
I.~Papadakis\altaffilmark{32,33},
L.~Pei\altaffilmark{10},
J.~S.~Schimoia\altaffilmark{1,34},
M.~Siegel\altaffilmark{11},
D.~Starkey\altaffilmark{6},
T.~Treu\altaffilmark{29,35,36},
P.~Uttley\altaffilmark{37},
S.~Vaughan\altaffilmark{21},
M.~Vestergaard\altaffilmark{38,39},
C.~Villforth\altaffilmark{6},
H.~Yan\altaffilmark{40},
S.~Young\altaffilmark{19},
and Y.~Zu\altaffilmark{1,41}
}

\altaffiltext{1}{Department of Astronomy, The Ohio State University,
  140 W 18th Ave, Columbus, OH 43210}
\altaffiltext{2}{Center for Cosmology and AstroParticle Physics, The
Ohio State University, 191 West Woodruff Ave, Columbus, OH 43210}
\altaffiltext{3}{Space Telescope Science Institute, 3700 San Martin
Drive, Baltimore, MD 21218}
\altaffiltext{4}{Department of Physics and Astronomy, The Johns
Hopkins University, Baltimore, MD 21218}
\altaffiltext{5}{Department of Physics and Astronomy, Georgia State
University, 25 Park Place, Suite 605, Atlanta, GA 30303}
\altaffiltext{6}{SUPA Physics and Astronomy, University of
St. Andrews, Fife, KY16 9SS Scotland, UK}
\altaffiltext{7}{Department of Physics, Western Michigan University,
1120 Everett Tower, Kalamazoo, MI 49008-5252} 
\altaffiltext{8}{School
of Physics and Astronomy, Raymond and Beverly Sackler Faculty of Exact
Sciences, Tel Aviv University, Tel Aviv 69978, Israel}
\altaffiltext{9}{Instituto de F\'{\i}sica y Astronom\'{\i}a, Facultad
de Ciencias, Universidad de Valpara\'{\i}so, Gran Bretana N 1111,
Playa Ancha, Valpara\'{\i}ıso, Chile}
\altaffiltext{10}{Department of Physics and Astronomy, 4129 Frederick
Reines Hall, University of California, Irvine, CA 92697}
\altaffiltext{11}{Department of Astronomy and Astrophysics, Eberly
College of Science, Penn State University, 525 Davey Laboratory,
University Park, PA 16802}
\altaffiltext{12}{Mullard Space Science Laboratory, University College
London, Holmbury St. Mary, Dorking, Surrey RH5 6NT, UK}
\altaffiltext{13}{Department of Statistics, The University of
Auckland, Private Bag 92019, Auckland 1142, New Zealand}
\altaffiltext{14}{Dipartimento di Fisica e Astronomia ``G. Galilei,''
Universit\`{a} di Padova, Vicolo dell'Osservatorio 3, I-35122 Padova,
Italy}
\altaffiltext{15}{INAF-Osservatorio Astronomico di Padova, Vicolo
dell'Osservatorio 5 I-35122, Padova, Italy}
\altaffiltext{17}{Department of Physics and Astronomy, Ohio
University, Athens, OH 45701}
\altaffiltext{18}{Department of Earth, Environment, and Physics, Worcester
State University, 486 Chandler Street, Worcester, MA 01602}
\altaffiltext{19}{Department of Astronomy, University of Maryland,
College Park, MD 20742-2421}
\altaffiltext{20}{University of Leicester, Department of Physics and Astronomy,
Leicester, LE1 7RH, UK}
\altaffiltext{21}{Astrophysics Science Division, NASA Goddard Space
Flight Center, Greenbelt, MD 20771}
\altaffiltext{22}{Spectral Sciences Inc., 4 Fourth Ave., Burlington,
MA 01803}
\altaffiltext{23}{Eureka Scientific Inc., 2452 Delmer St. Suite 100,
Oakland, CA 94602}
\altaffiltext{24}{Space Science Center, Morehead State University, 235
Martindale Dr., Morehead, KY 40351}
\altaffiltext{25}{Department of Physics and Astronomy, York
University, Toronto, ON M3J 1P3, Canada }
\altaffiltext{26}{SRON Netherlands Institute for Space Research,
Sorbonnelaan 2, 3584 CA Utrecht, The Netherlands}
\altaffiltext{27}{Department of Physics and Astronomy, Univeristeit
Utrecht, P.O. Box 80000, 3508 Utrecht, The Netherlands}
\altaffiltext{28}{Leiden Observatory, Leiden University, PO Box 9513,
2300 RA Leiden, The Netherlands}
\altaffiltext{29}{Department of Physics, University of California,
Santa Barbara, CA 93106}
\altaffiltext{30}{Departamento de Astronomia, Universidad de Chile,
Camino del Observatorio 1515, Santiago, Chile}
\altaffiltext{31}{University of Southampton, Highfield, Southampton,
SO17 1BJ, UK}
\altaffiltext{32}{Department of Physics and Institute of Theoretical
and Computational Physics, University of Crete, GR-71003 Heraklion,
Greece}
\altaffiltext{33}{IESL, Foundation for Research and Technology,
GR-71110 Heraklion, Greece}
\altaffiltext{34}{Instituto de F\'{\i}sica, Universidade Federal do
Rio do Sul, Campus do Vale, Porto Alegre, Brazil}
\altaffiltext{35}{Department of Physics and Astronomy, University of
California, Los Angeles, CA 90095-1547}
\altaffiltext{37}{Astronomical Institute `Anton Pannekoek,' University
of Amsterdam, Postbus 94249, NL-1090 GE Amsterdam, The Netherlands}
\altaffiltext{38}{Dark Cosmology Centre, Niels Bohr Institute,
University of Copenhagen, Juliane Maries Vej 30, DK-2100 Copenhagen,
Denmark}
\altaffiltext{39}{Steward Observatory, University of Arizona, 933
North Cherry Avenue, Tucson, AZ 85721}
\altaffiltext{40}{Department of Physics and Astronomy, University of
Missouri, Columbia, MO 65211}
\altaffiltext{41}{Department of Physics, Carnegie Mellon University,
5000 Forbes Avenue, Pittsburgh, PA 15213}

 \footnotetext[16]{NSF Postdoctoral Research Fellow}
\footnotetext[36]{Packard Fellow}

\begin{abstract} 
We describe the first results from a six-month long
reverberation-mapping experiment in the ultraviolet based on 171
observations of the Seyfert 1 galaxy NGC~5548 with the Cosmic
Origins Spectrograph on the {\em Hubble Space Telescope}.
Significant correlated variability is found in the continuum and
broad emission lines, with amplitudes ranging from $\sim30$\% to a
factor of two in the emission lines and a factor of three in the
continuum.  The variations of all the strong emission lines lag
behind those of the continuum, with \heii\,$\lambda1640$ lagging
behind the continuum by $\sim2.5$\,days and \Lya\,$\lambda1215$,
\civ\,$\lambda1550$, and \siiv\,$\lambda1400$ lagging by $\sim
5$--6\,days.  The relationship between the continuum and emission
lines is complex. In particular, during the second half of the
campaign, all emission-line lags increased by a factor of 1.3--2 and
differences appear in the detailed structure of the continuum and
emission-line light curves.  Velocity-resolved cross-correlation
analysis shows coherent structure in lag versus line-of-sight
velocity for the emission lines; the high-velocity wings of \civ\
respond to continuum variations more rapidly than the line core,
probably indicating higher velocity BLR clouds at smaller distances
from the central engine. The velocity-dependent response of \Lya,
however, is more complex and will require further analysis.
\end{abstract}



\section{INTRODUCTION}
\label{section:intro}
\subsection{The Broad-Line Region}
\label{section:introBLR}
One of the most prominent characteristics of the ultraviolet (UV),
optical, and near-infrared (NIR) spectra of active galactic nuclei
(AGNs) is the presence of broad emission lines. While we know that
these features arise on scales not much larger than the accretion
disk, their physical nature remains one of the major unsolved
mysteries in AGN astrophysics. A particularly important feature of the
broad emission lines is that they are, by definition, resolved in
line-of-sight (LOS) velocity, and their large widths leave little
doubt that the primary broadening mechanism is differential Doppler
shifts due to the motion of individual gas clouds, filaments, or
more-or-less continuous flows around the central black hole. 

However, it is not possible to establish the broad-line region (BLR) kinematics
simply by inverting the line profiles because this inverse problem is
degenerate, with a wide variety of simple velocity models providing
satisfactory fits \citep[e.g.,][]{Capriotti80}. The existing evidence on the 
BLR kinematics is
ambiguous: some of this gas may flow inward, helping to feed the
central black hole. Extended, flattened, rotating
disk-like structures seem to be important in at least some BLRs, as
shown statistically for radio-loud AGNs
\citep{Wills86,Vestergaard00,Jarvis06}, by the pronounced double-peaked
profiles observed in some sources \citep[e.g.,][]{Eracleous94,Eracleous03,
Strateva03,Gezari07,Lewis10}, and from
spectropolarimetry \citep{Smith04,Young07}.  There is evidence of 
the importance of the black
hole gravity in dominating the motion of the BLR gas \citep{Peterson04},
although radiation pressure may also play a role \citep{Marconi08,Netzer10}.

On the other hand,
developments over the last two decades re-open the interesting
possibility that much of the emitting BLR gas is due to outflowing
winds \citep[e.g.,][]{Bottorff97,Murray97,Proga00,Everett03,Elvis04,Young07},
perhaps connected to the
outflows detected in absorption features 
\citep[e.g.,][]{Hamann04,Krongold05,Krongold07,Kriss11,Kaastra14,Scott14}, whose
kinematics and energetics are also poorly understood. 
The unknown dynamics of the BLR gas represents a serious gap in our
understanding of AGNs and in the calibrations needed for the study of
black-hole/host-galaxy co-evolution up to very high redshifts.

There have been many attempts to model the physics of the BLR. In general,
photoionization equilibrium models can reproduce the line intensities,
but self-consistent models that provide simultaneous solutions to the
line intensities, profiles, and variability are lacking. The locally
optimally emitting cloud model \citep{Baldwin95,Korista00} and the
stratified cloud model \citep{Kaspi99} explain most observed line
intensities and some of the observed time lags between the continuum
and emission lines. However, they lack the important kinematic
ingredients required to explain the observed line profiles. 

\subsection{Reverberation Mapping}
\label{section:introRM}  
In order to understand the structure and kinematics of the BLR, we
must break the degeneracy that comes from the study of the line
profiles alone. We can do this by using reverberation mapping (RM) to
determine how gas at various LOS velocities responds to continuum
variations as a function of light travel-time delay
\citep{Blandford82,Peterson93,Peterson14}.

Over the last quarter century, the RM technique has become
a standard tool for investigating the BLR.  In its simplest form, RM
is used to determine the mean time delay between continuum and
emission-line variations, typically by cross-correlation of the
respective light curves. It is assumed that this represents the mean
light-travel time across the BLR. By combining this with the
emission-line width, which is assumed to reflect the velocity
dispersion of gas whose motions are dominated by the mass of the
central black hole, the black hole mass can be estimated. RM in this
form has been used to measure the black hole masses in over 50 AGNs
\citep[for a recent compilation, see][]{Bentz15} to a typical accuracy
of $\sim 0.3$\,dex. Important findings that have arisen from
these RM studies include the following:
\begin{enumerate}
\item In a given AGN, emission lines that are characteristic of
higher-ionization gas respond more rapidly to continuum
flux variations than those characteristic
of lower-ionization gas, indicating ionization stratification within
the BLR \citep{Clavel91,Reichert94}.
\item There is an inverse correlation between the time delay, or lag
$\tau$, for a particular emission line and the Doppler width $\Delta
V$ of that emission line. The relationship for a given AGN is
consistent with the virial prediction $\Delta V \propto \tau^{-1/2}$
\citep{PW99,PW00,Kollatschny03,Peterson04,Bentz10a}. Without this
relationship, RM masses would be highly dubious.
\item There is an empirical relationship between the AGN luminosity
$L$ and the radius of the BLR $R$ (hereafter the $R$--$L$
relationship) that is well-established only for the \Hbeta\ emission
line \citep{Kaspi00,Kaspi05,Bentz06,Bentz09,Bentz13}.  Limited data on
\civ\,$\lambda1549$ indicates a similar relationship applies to
that line as well \citep{Peterson05,Vestergaard06,Kaspi07,Park13}.
The existence of $R$--$L$ relationships for both low-ionization and
high-ionization lines has been independently confirmed by 
gravitational microlensing observations \citep{Guerras13}.
\end{enumerate} 
The $R$--$L$ relationship is of particular interest as
it allows estimation of the central black-hole mass based on a single
spectrum from which the line width is measured and the BLR radius is
inferred from the AGN luminosity. This neatly bypasses the need for a
direct RM measurement of the emission-line time lag. RM is necessarily
resource intensive: even to determine the mean time delay for an
emission line typically requires some 30--50 well-spaced high-quality
spectrophotometric observations or a good measure of luck for fewer
observations.  The $R$--$L$ relationship is very important as the
RM-based mass determinations anchor empirical scaling relationships
\citep[e.g.,][]{McLure02,Vestergaard02,Shields03,Grupe04,
Vestergaard04,Greene05,Mathur05,
Kollmeier06,Vestergaard06,Salviander07,Treu07,McGill08,Park13,Park15,Netzer14} 
that are used to estimate the masses of
quasar black holes in large numbers
\citep[e.g.,][]{Vestergaard08,Vestergaard09,Shen11,DeRosa14}.
Virtually all quasar mass
estimates and their astrophysical uses are tied to RM.

Measurement of the mean lag and line width for a given emission line
provides important, though limited, information about the BLR and the
central mass of the AGN.  We are only now beginning to
realize the full power of RM through velocity-resolved investigations
of the BLR response.  The first generation of successful RM programs
provided sufficient understanding of AGN variability and BLR response
times to design programs that could effectively extract
velocity--dependent information that would lead to an understanding of
the structure and kinematics of the BLR through recovery of
``velocity--delay'' maps from RM data \citep{Horne04}.  The
relationship between the continuum variations $\Delta C(t)$ and
velocity-resolved emission-line variations $\Delta L(V,t)$ is usually
described as
\begin{equation}
\label{eq:trans} \Delta L(V,t) = \int_{0}^{\infty} \Psi (V,\tau)
\Delta C(t-\tau )d\tau,
\end{equation} where $\Psi(V,\tau)$ is the ``response function,''
or velocity--delay map \citep{Horne04}. As can be
seen by inspection, $\Psi(V,\tau)$ is simply the observed
emission-line response to a delta-function continuum outburst.  The
velocity--delay map is simply the BLR geometry and kinematics
projected into the two observable quantities of LOS velocity and time
delay relative to the continuum. This linearized echo model is
justified by the fact that the continuum and emission-line variations
are generally quite small (10--20\%) on reverberation time scales
\citep[see also][]{Cackett06}.  The technical goal of a reverberation
program such as the one described here is to recover the velocity--delay map $\Psi(V,\tau)$ from the
data and thus infer the geometry and kinematics of the BLR.

Time-resolved velocity--delay maps have now been obtained for a
handful of AGNs \citep[e.g.,][]{Bentz10b,Brewer11,Pancoast12,Grier13,Pancoast14}, but
only for optical lines (the Balmer lines, \hei\,$\lambda5876$, and
\heii\,$\lambda4686$). In general, these suggest flattened geometries
at small to modest inclinations and some combination of 
virialized motion and infall. An outflow signature has been observed in
only one case, NGC\,3227 \citep{Denney09}.

The lack of velocity--delay maps for UV lines,
on the other hand, leaves us with a very incomplete understanding of
the BLR. It is, in fact, the high-ionization level UV resonance lines (e.g.,
\civ\,$\lambda1549$, \siiv$\lambda1400$, \Lya\,$\lambda1215$) that
might be expected to dominate any outflowing component of the
BLR. The optical lines, in contrast, generally seem to arise in
disk-like structures with infall components \citep[e.g.,][]{Pancoast14}.

RM studies in the UV have been limited. Several observing campaigns
were undertaken with the {\em International Ultraviolet Explorer
(IUE)} or {\em Hubble Space Telescope (HST)} or both on (i) NGC\,5548
\citep{Clavel91,Korista95}, (ii) NGC\,3783 \citep{Reichert94}, (iii) Fairall\,9
\citep{Clavel89,Rodriguez97}, (iv) 3C\,390.3 \citep{Obrien98}, (v) NGC 7469
\citep{Wanders97}, (vi) NGC\,4151 \citep{Clavel90,Ulrich96,Crenshaw96},
(vii) Akn\,564 \citep{Collier01}, and (viii) NGC\,4395 \citep{Peterson05}.  With
the exception of Akn\,564, which showed essentially no emission-line
variability over a comparatively short campaign, all of these programs
yielded emission-line lags, but only limited information about the
detailed response of the UV emission lines
\citep[e.g.,][]{Horne91,Krolik91,Wanders95,Done96}.  The existing
velocity-delay map for NGC\,4151 shows some incipient structure in
\civ\,$\lambda1549$ and \heii\,$\lambda1640$ and a general shape that
seems to be consistent with a virialized BLR \citep{Ulrich96}.

\subsection{The AGN STORM Project}
\label{section:introSTORM}
Given the importance of the UV emission lines in the photoionization
equilibrium of the BLR gas and the probable differences between the
geometry and kinematics of the high and low-ionization gas in the BLR,
we have undertaken a large RM program in the UV using the Cosmic
Origins Spectrograph (COS; \citealt{Green12}) on \HST\ (\HST\ Program GO-13330),
the AGN Space Telescope and Optical Reverberation Mapping (AGN STORM)
Project, in the first half of 2014.  The program was designed with certain
specific goals in mind:
\begin{enumerate}
\item Determine the structure and kinematics of the high-ionization
BLR through observations of the variations in the \civ\,$\lambda1549$, \Lya\,$\lambda1215$,
\nv\,$\lambda1240$, \siiv\,$\lambda1400$, and \heii\,$\lambda1640$
emission lines.\footnote{We note
that three of these lines are actually doublets:
\nv\,$\lambda\lambda1239, 1243$, \siiv\,$\lambda1394, 1403$, and
\civ\,$\lambda\lambda1548, 1551$. Moreover, the
\heii\ feature is blended with \oiii]\,$\lambda\lambda1661$, 1665 and
\siiv\ is blended with the quintuplet
\oiv]\,$\lambda\lambda1397.2$, 1399.8, 1401.2, 1404.8, 1407.4,
where the second, third, and fifth transitions dominate.} 
\item Carry out simultaneous ground-based observations of (a) the
high-ionization optical line \heii\,$\lambda4686$ for direct
comparison with \heii$\,\lambda1640$ and (b) the Balmer lines,
particularly \Hbeta\,$\lambda4861$, to determine the structure and
kinematics of the low-ionization BLR.  Although the optical spectrum
is extremely well-studied \citep[][and references therein]{Peterson02,
Bentz07,Bentz10a,Denney10}, simultaneous
observations are necessary, as the dynamical timescale for the BLR in
NGC\,5548 is only a few years.
\item Compare in detail the continuum variations in the UV (at
$\sim1350$\,\AA) with those at other wavelengths
\citep[see][hereafter Paper II]{Edelson15} and
infer the structure of the continuum-emitting region.
\end{enumerate} 
The motivation for the UV/optical continuum comparison
is multifold:
\begin{enumerate}
\item Delays between continuum variations at longer versus shorter
wavelengths have been detected or hinted at in a number of sources
\citep[e.g.,][]{Wanders97,Collier98,Collier01,Peterson98,Sergeev05,Cackett07,
McHardy14,Shappee14}. Such delays can provide insight into the structure,
geometry, and physics of the continuum-emitting region.
\item Velocity--delay maps recovered using the UV continuum as the
driving light curve (Equation \ref{eq:trans}) are expected to be of higher
fidelity than those obtained from the optical continuum because the
observable UV is closer in wavelength to the ionizing continuum
($\lambda < 912$\,\AA) that powers the emission lines. The optical
continuum is not only a slightly time-delayed version of the UV
continuum, but it seems smoothed somewhat as well
\citep{Shappee14,PetersonPlus14}, which might make it difficult to recover
detailed structure in the velocity--delay maps.
\end{enumerate}

Our \HST\ program afforded a valuable opportunity for exploring AGN
behavior at high time resolution for an extended period at wavelengths
beyond those covered by our  \HST\ COS spectra. The \HST\
program is the anchor of a much broader AGN STORM project to address
broader issues through observations across the electromagnetic
spectrum. This paper serves as the first in a series.

Of special interest is the possibility of using short-timescale
lags between variations in different continuum bands to map the temperature
structure of the accretion disk. The {\em Swift} satellite \citep{Gehrels04}
is especially suitable for such a study because of its broad wavelength coverage
(hard X-ray through $V$-band) and ability to execute high-cadence observations over
an extended period of time. In Paper II, we present the results of a
four-month program of high-cadence (approximately twice per day)
multiwavelength observations with {\em Swift}. Additional papers in this 
series will describe high-cadence ground-based photometry from
the near UV through the NIR. We will also present results from
a program of ground-based spectroscopy that is similar in cadence to the \HST\
COS observations, but covers a somewhat longer temporal baseline. Other additional
papers will present results on the variable absorption features 
and on our efforts to decipher the broad emission-line variations and determine
the structure and geometry of the BLR.

In Section 2, we describe the observations and data processing,
including a discussion of the program design and 
a complete description of how the standard data reduction
pipeline was modified to meet our stringent calibration
requirements. We describe our initial data analysis 
and results in Section 3, and in Section 4, we briefly discuss
the first results from our program and place these results in the context
of previous monitoring campaigns on NGC 5548.
When necessary, we assume a $\Lambda$CDM cosmology with $H_0 =
70\,{\rm km}^{-1}\,{\rm s}^{-1}\,{\rm Mpc}^{-1}$, $\Omega_{\rm M} =
0.28$, and $\Omega_{\Lambda} = 0.72$ \citep{Komatsu11}.

\section{Observations and Data Reduction}

\subsection{Program Design} 
\label{sec:design}

RM is a resource-intensive activity that requires obtaining high
signal-to-noise ratio ($S/N$) homogeneous spectra at sufficiently high
spectral resolution to resolve the gross kinematics of the BLR.
Spectra must be obtained at a high cadence over a temporal baseline
that is longer than the typical variability timescale of the AGN.
Given the inherent risks of RM programs due to the unpredictability of
AGN variability, it is essential that our experimental design assures
a successful outcome, yet is as economical with observing time as
possible. The first consideration is that each epoch of observation
should require no more than one \HST\ orbit per ``visit'' which
restricts the integration time per visit to $\sim45-50$ minutes. This
consideration limits us to relatively bright nearby Seyfert 1
galaxies. COS is clearly the instrument of choice for such a project,
as it is a very sensitive, high spectral resolution spectrometer. Its
native resolution ($R > 20000$) is high enough to allow us to trade
off resolution and $S/N$ in the data processing phase. In order to
schedule the observatory efficiently, a cadence of one visit per day
or longer is required.

We therefore want to target an AGN that has 
a \civ-emitting region several light days in extent, and this requires
a source with $\log L_{\lambda}(1350\,{\rm \AA}) / ({\rm ergs\, s}^{-1}) \gtsim 43.5$ 
\citep[e.g.,][]{Kaspi07}. This led us immediately to select as a
target the well-studied Seyfert 1 galaxy NGC\,5548 ($z = 0.017175$).
NGC\,5548 is probably the best-studied AGN by RM, with historical
optical spectroscopy extending as far back as the early 1970s
\citep{Sergeev07}. Importantly, it has never been known to go into a
``dormant state,'' as observed recently in the case of Mrk 590 
\citep{Denney14}, that
would preclude a successful reverberation campaign and, historically,
self-absorption in the UV resonance lines has been minimal
\citep{Crenshaw99}, although strong absorption appeared in
2013 \citep{Kaastra14}.

The remaining adjustable parameter is the duration of the campaign. We
investigated this using Monte Carlo simulations similar to those
described by \cite{Horne04}.  Using recent developments in
statistically modeling AGN light curves \citep{Kelly09,Kozlowski10,MacLeod10}, we can make
very robust models of the expected continuum behavior of NGC
5548. Quasar light curves are well-described by a stochastic process,
the damped random walk.  The process is described by an amplitude
$\sigma$ and a damping timescale $\tau_{\rm d}$, which for NGC 5548 in
the optical are measured to be $\sigma = 0.89 ^{+0.30}_{-0.20} \times
10^{-15}\ {\rm ergs\ s}^{-1}\, {\rm cm}^{-2}\,{\rm \AA}^{-1}$ and
$\tau_{\rm d} = 77^{+59}_{-34}$ days, respectively
\citep{Zu11}.  We used these measured properties of NGC 5548 to
simulate the continuum variations; this is a conservative choice as
the UV continuum can be expected to show both higher amplitude and
shorter time-scale variations, both of which are an advantage. We then
convolved the artificial light curves with model velocity--delay maps
for several lines to provide an artificial spectrum.  As described by
\cite{Horne04}, we adopted a BLR model with an extremely challenging
velocity--delay map for these simulations, a Keplerian disk with a
single two-armed spiral density wave.  While this is unlikely to be the actual
AGN BLR geometry, it provides a challenging
test: if we can recover such a complex velocity--delay map correctly,
then we can certainly hope to recover others of comparable complexity
and would have no difficulties with geometries like those that 
have been recovered for optical lines
\citep[e.g.,][]{Bentz10b,Pancoast12,Grier13,Pancoast14}.

We modeled the emissivity and response of each line realistically
using a grid of photoionization equilibrium models \citep{Horne04}. We sampled
the artificial spectra to match our proposed observations, including
noise. We then modeled the artificial spectra to recover the
velocity--delay maps using {\tt MEMECHO} \citep{Horne94,Horne04}.  Simulations
based on characteristics of previous RM experiments yield
velocity--delay maps with noise levels similar to those obtained from
the actual data, demonstrating the verisimilitude of our simulations
\citep[see][for examples]{Horne04}.

The goal of our simulations was to determine the minimum duration
program that would allow us to recover a velocity--delay map with a
high probability of success.  For COS-like observations (in terms of
$S/N$ per visit and spectral resolution), our initial simulations
indicated that reliable velocity--delay map recovery for a strong line
(e.g., \civ) required between 130 to 200 days. A finer grid of models
showed that in 10 of 10 simulations, a high-fidelity velocity--delay
map was recovered after 180 days, which was thus adopted as the
program goal. A much longer program at this sampling rate would in
any case be precluded by the accessibility of the target to \HST.

Because of the long duration of the proposed program, we also
considered the possible impact of losses of data due to instrument or
spacecraft safing events. Short safing events occur frequently enough
that we needed to assess their impact. Based on the record for \HST\
and COS in Cycles 17--20,
there might be two spacecraft events that lose 2--3 days each and one
COS event that loses 2 days
over a stretch of 180 consecutive days. By repeating a subset of our simulations,
we found that losses of such small numbers of observations would have
no impact on our ability to recover the velocity--delay maps.  The
simulations also allowed us to assess the impact of early termination
of our experiment due to a major failure. If a program was terminated
at $\sim100$ days, the probability that the data would yield a useful
(but not a detailed) velocity--delay map would be $\sim
50$\%. However, a program as short as 75 days would have a very low
probability ($\sim10$\%) of success.

The key to a successful RM campaign is that it must be long enough
that favorable continuum variability characteristics 
become highly probable. That this is essentially guaranteed to happen during a
180-day experiment played a major role in selecting NGC 5548 as our
target.

\subsection{COS Observations}
\label{sec_COS_Observations}

Observations were made in single-orbit \HST\ COS visits approximately daily from 
2014 February 1 through July 27. Of the 179
scheduled visits, 171 observations were executed successfully and 8
were lost to safing events or target acquisition failures (very close
to the expected number of losses).

In each visit, we used the G130M and G160M gratings to observe the UV
spectrum over the range 1153--1796\,\AA\ in four separate
exposures. Exposure times were selected to provide $S/N \gtsim 100$ 
when measured over velocity bins of $\sim500\,\kms$. During each
visit, we obtained two 200-second exposures with G130M centered at
1291\,\AA\ and 1327\,\AA\ and two 590-second exposures with G160M
centered at 1600\,\AA\ and 1623\,\AA.

The COS far-ultraviolet detector is a windowless, crossed delay-line
microchannel plate stack that is subject to long-term charge depletion.
To extend the useful lifetime of the detector, we positioned the
spectrum so that bright geocoronal airglow lines (e.g., Ly$\alpha\lambda1215$)
and AGN emission lines (e.g., redshifted Ly$\alpha$) 
would not always fall on the same area of the detector.  
First, we alternated the target acquisition between the G130M/1291 and
the G130M/1327 configurations. The G130M/1327 configuration is then
followed by a G130M/1327/FP-POS=3 exposure\footnote{FP-POS values
refer to small displacements of the spectrum on the detector 
in the dispersion direction in order
to minimize the effects of fixed-pattern noise.}, and by a G130M/1291
exposure alternating among FP-POS=1, 2, and 4. The G130M/1291 configuration
is instead followed by a G130M/1291/FP-POS=3 exposure, and by a
G130M/1327 exposure alternating among FP-POS=1, 2, and 4.
Second, we alternated the FP-POS for the G160M/1623 exposure between
FP-POS=1 and FP-POS=2. We could not vary the settings for the G160M/1600 and
G160M/1623 further because we needed to ensure the coverage of the entire
wavelength range while keeping the detector gap from falling on
the redshifted C\,{\sc iv}\,$\lambda1549$ emission line.

Finally, we used four additional orbits to improve our
understanding of the COS flux calibrations (see Section
\ref{sec_Data_Reduction}).  During these additional visits,
we observed two of
the standard stars (WD 0308--565 and WD 1057+719) employed to obtain
sensitivity functions \citep{Massa2014} at the same detector locations we used for the
reverberation program.  The observations were taken using all the
instrument configurations employed in our primary observing program.

\subsection{Data Reduction}
\label{sec_Data_Reduction}
We used the {\tt CalCOS} pipeline v2.21 for the bulk of our data
processing.  The absolute flux calibration of the COS reduction pipeline
is reported to be accurate to $\sim5\%$ and the relative flux calibration
is good to better than $\sim2\%$ \citep{Holland2014}. We are primarily interested in the quality
of the relative flux calibration as we are looking for very small-scale variations 
on short timescales; we need the fluxes to be stable and
repeatable across the spectrum.  We found, however, that there were local 
variations in the precision of the fluxes that necessitated 
improvements.

To produce a final dataset with a flux calibration that is
everywhere precise at the 2\% level, we refined the existing
calibration reference files and applied a post-{\tt CalCOS} pipeline
to further process the data. The main areas of improvement include
refinements to the dispersion solution, fixed-pattern noise mitigation, 
the sensitivity function, and the time dependent sensitivity (TDS)
functions, as
outlined below. The final data product consists of
one combined spectrum per grating per day. Airglow emission lines 
(\oi\,$\lambda\lambda 1302.2,1306$ and \Ni\,$\lambda\lambda 1199.5,1200.7$) were filtered
from the data by removing events detected when \HST\ was in daylight. The
spectra were further binned by 4 pixels in order to increase the
$S/N$ per spectral element of the AGN continuum.
This binning still results in two binned pixels per COS resolution element.

\subsubsection{Dispersion solution}
The COS wavelength solution has a quoted uncertainty of 
$\sim15$\,km\,s$^{-1}$ \citep{Holland2014}.  As the STIS uncertainty is
$<5$\,km\,s$^{-1}$ \citep{Hernandez2014}, we refined the dispersion
solutions for our COS dataset using previous observations of NGC~5548
taken with the STIS E140M/1425 mode in 1998 (PID 7572, PI: Kraemer). To accomplish
this, we cross-correlated the line profiles of strong interstellar
medium absorption features between each COS observation and the STIS reference
spectrum. We used 19 interstellar absorption features, ranging from
\Sizw\,$\lambda1190$ at the short-wavelength end to
\alii\,$\lambda1670$ at the long-wavelength end.
A linear correction to the initial wavelength solution was
then computed across each detector segment and applied directly to the
extracted spectra. With this correction, measurements of the root mean square 
(rms) of the residual offsets decreased from $\sim15$\,km\,s$^{-1}$ to 
$<6$\,km\,s$^{-1}$.

\subsubsection{Fixed-pattern noise}
The standard reference files used in the {\tt CalCOS} pipeline correct
for only the most prominent fixed-pattern noise features such as the
quantum-efficiency gridwires, low-order response variations, and large
geometric distortion artifacts \citep{Ely2011}.  Usually, users 
combine multiple FP-POS positions to smooth over the remaining
features. However, this was not possible for our dataset, since only a
single FP-POS setting was used for each central wavelength setting
in each orbit (see Section \ref{sec_COS_Observations}).

To correct these features to a higher degree we derived one-dimensional
pixel-to-pixel flats (``p-flats'').  These flats were produced by
combining normalized, high signal-to-noise ratio white dwarf spectra in
detector space, following the method described by \citet{Ely2011}.
The white dwarf spectra used were taken as part of the {\it HST}/CAL
program 12806 (PI: Maasa) and used the same detector locations as used for the NGC~5548
datasets.

To test the effects of our p-flat correction, we combined the 171
spectra reduced both with and without the application of the
p-flats. The {\it S/N} per pixel in 5\,\AA\ continuum regions increased
from $\sim75$ to $\sim80$ for the G130M grating, and from $\sim60$ to
$\sim$$100$ for the G160M grating through the removal of small localized
flux calibration errors by the p-flat correction.
The improvement for the G130M combined
spectrum is less dramatic because (a) we rotated among the four FP-POS
settings,  and (b) the G130M grating disperses more widely in the
cross-dispersion direction, and thus intrinsically averages the 
fixed-pattern noise
over a larger area of the detector.

\subsubsection{Sensitivity functions and TDS}
The COS flux calibration is done in two steps: (a) derivation of static
sensitivity functions and (b) characterization of the time evolution of
the sensitivity through the TDS correction \citep{Holland2014}.  Thanks to
the existing calibration program that monitors the TDS variations (PID
13520), we had bi-monthly observations of the standard star
WD~0308--565 for 3 out of the 4 central wavelength settings we are
using in our program (G130M/1291, G130M/1327 and G160M/1623). Standard
star data were obtained in 2014 February, April, June, and August. By
analyzing these calibration data, together with the data collected
during our additional calibration orbits (Section \ref{sec_COS_Observations}),
we verified that both the static and time-dependent response functions
vary more with instrument configuration than currently modeled by the 
{\tt CalCOS} pipeline.
While {\tt CalCOS} assumes that both the sensitivity
function and the TDS correction vary only as a function of wavelength,
we were able to improve the relative flux calibrations and reach our
required level of precision by (a) obtaining sensitivity functions
individually for each configuration (one function per wavelength setting per
FP-POS per detector segment), and (b) computing the TDS correction
individually for each wavelength setting observed as part of the routine
calibration program.

We estimate the quality of the flux calibration by inspection of the fractional
residuals $f_{\rm res}$ of the calibrated standard star spectra and their respective
{\tt CalSPEC} stellar model (the same models employed by the standard
pipeline reference file),
\begin{equation}
f_{\rm res} = \frac{f_{\rm WD,obs} - f_{\rm Model}}{f_{\rm Model}},
\end{equation}
where both $f_{\rm WD,obs}$ and $f_{\rm Model}$ are binned over 1\,\AA\
using a boxcar filter in order to increase the {\it S/N} per spectral
element. While the visual inspection of the residuals as a function of
wavelength allows us to identify and correct for local biases, we use
the mean value of the distribution of the residuals as an indicator of
a global bias in the calibration. 
The flux calibration uncertainty is an estimate of the limit of the stability of the flux
calibration at a given time. However, since the overall instrument
sensitivity evolves with time, and our final spectra are obtained from
the combination of multiple settings for each grating, we
conservatively define the 
fractional precision error $\delta_{\rm P}$ for each
grating as the maximum fractional
uncertainty computed for any of the
wavelength settings.

The new sensitivity functions were derived from
spectra of the standard star WD~0308--565 for the G130M settings, and
of WD~1057+719 for the G160M settings (PID 12806).
While one individual sensitivity function per detector
segment characterizes the full grating (data from different settings
are averaged together) in {\tt CalCOS}, we built one independent sensitivity
function for each wavelength setting and FP-POS used in our program.

By comparing the bi-monthly WD~0308--565 data, we found that residuals 
with respect to the stellar models were greatly reduced if the TDS
corrections were computed individually for each of the 
wavelength settings (G130M/1291, G130M/1327 and G160M/1623), instead of averaging
the data over multiple modes. Additional improvements were obtained by
increasing the number of time intervals over which the TDS trends are
computed and by redefining the wavelength ranges used in the
analysis. Unfortunately, there are insufficient calibration data for
the longest wavelengths in the G160M spectra, so we were forced to truncate
these spectra at 1750\,\AA.  

In spite of these
improvements, the available data did not allow us to conduct any tests
on the remaining setting (G160M/1600). This setting is particularly
important for our scientific goals since it includes most of the broad
\civ \ emission line (section \ref{section:introSTORM}). Moreover, since all
the calibration data for TDS monitoring purposes are obtained only
using FP-POS=3, they did not allow us to test for any residual
dependence of the TDS correction on FP-POS configuration. These are
the two reasons that motivated us to request further calibration
data (see Section \ref{sec_COS_Observations}). These data for WD~0308--565
and WD~1057+719 collected in 2014 September allowed us to derive an
independent set of sensitivity functions. By comparing the new sensitivity
functions with the originals, appropriately corrected for time
evolution of the TDS, we were able: 
\begin{enumerate}
\item{To identify the best possible TDS correction attainable for the
G160M/1600 setting with the current TDS calibration data. The
current {\tt CalCOS} TDS correction for this configuration was
obtained from combining both G160M/1577 and G160M/1623
    data. Although this correction is not ideal,
it minimizes both the global bias and the flux
    calibration uncertainty when compared to the TDS corrections
    obtained individually from either the G160M/1577 or the G160M/1623
    settings.}
\item{To verify that the TDS correction does not vary strongly with
    FP-POS settings. While comparing the 
   residuals with respect to the stellar models 
 shows structure unique to each FP-POS setting, the level of the
    local biases is such that both the global bias and
    the flux calibration uncertainty 
can be considered stable for each wavelength setting (the maximum
    deviation in the width of the residual distribution is $\sim0.2\%$).}
\end{enumerate}

With the new flux calibration and TDS characterization, our global biases
are consistent with zero and the overall 
fractional precision is $\delta_{\rm P}
\sim1.1$\% and $\sim1.4\%$, respectively, for the G130M and G160M
settings, compared to $\delta_{\rm P} \sim1.4\% {\rm\ and} \sim3.5\%$
for the standard pipeline. We intend to make these improvements available
to other COS users.

\subsubsection{Sensitivity offsets for the final week of data}
During the final week of the observing campaign, the operating high
voltage (HV) for one of the COS detector segments was increased to
combat the negative effects of ``gain-sag'' \citep{Sahnow2011}.
While necessary to provide well-calibrated data, this HV change also
has the effect of introducing small changes in the detector response.
Using WD~0308--565 observations from our calibration orbits
taken at the same HV as the rest of the campaign and contemporaneous
TDS monitoring observations of the same calibration target (taken at the increased
HV), we were able to estimate a HV bias correction for the G130M
grating from a direct comparison of the spectra.  This bias was
measured to be 1\% without any detectable dependence on wavelength or
cenwave setting.  Unfortunately, the same procedure could not be done
for the G160M grating as our additional orbits and the TDS
observations used different standard stars (WD~0308--565 instead of
WD~1057+719). Instead, the bias estimate for this grating was obtained
by analyzing the time evolution of the mean flux in overlapping
regions of G130M/1327 at the lower HV setting and G160M/1600 at the higher 
HV setting.
This analysis gave a plausible estimate of a 1\% bias. However, with
such a limited amount of data at the lower HV and a narrow overlapping
wavelength range, the estimate lacks the accuracy of the G130M bias estimate.

\section{Data Analysis}
\label{sec_Data_analysis}
\subsection{Mean and RMS Spectra}
\label{sec:MeanRMS}
For an initial look at the spectral variations, we define  
G130M and G160M mean spectra as
\begin{equation}
\label{eq:meanspectrum}
\overline{F}\left(\lambda \right) 
 = \frac{1}{N} \sum_{i=1}^{N} {F_i\left(\lambda \right)} \ ,
\end{equation}
where $F_i$ is the $i$th spectrum of the series of $N=171$ spectra.
Similarly, the rms residual spectrum (hereafter referred to simply
as the RMS spectrum) is defined as 
\begin{equation}
\label{eq:rmsspectrum}
S \left(\lambda \right) = \left\{ {\frac{1}{N-1} \sum_{i=1}^{N} {\left[ F_i\left(\lambda \right) -  \overline{F}\left(\lambda \right) \right]^2 }}\right\}^{1/2} \ .
\end{equation}
The RMS spectrum is especially useful as it isolates the variable part of the spectrum;
constant components disappear, though sometimes small residuals are visible
in the case of strong features.

The  statistical uncertainty in the mean spectra is
\begin{equation}
\label{eq:staterror}
\sigma_{\overline{F}} \left( \lambda \right) = \frac{1}{N} 
\left\{ \sum_{i=1}^{N}  \sigma^2_{F_i}\left( \lambda \right)  \right\}^{1/2},
\end{equation}
where $\sigma_{F_i}$ is the error spectrum of the $i$th spectrum in the series.

\begin{figure*}
\centering
\includegraphics[width=0.9\textwidth]{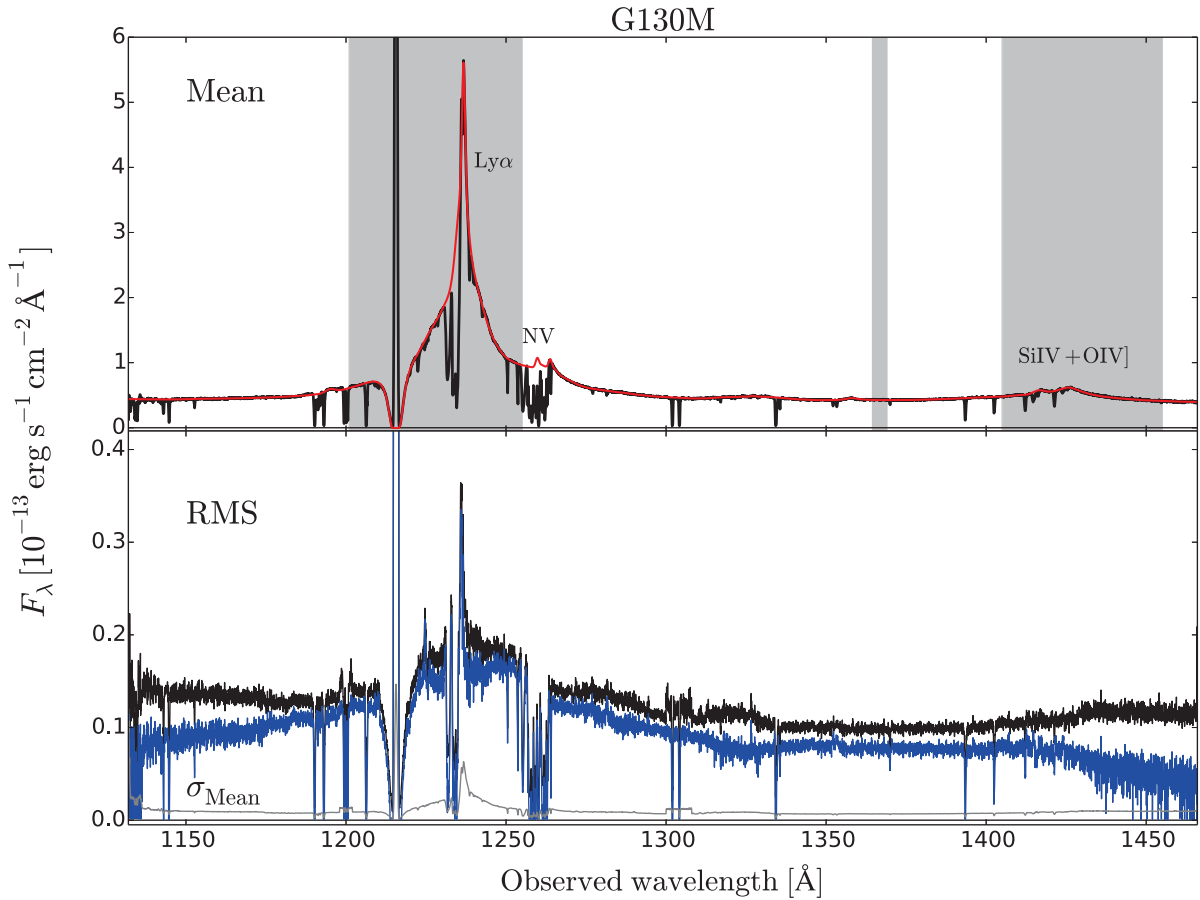}
\caption{Spectra obtained with the G130M grating. The shaded areas show the integration regions
defined in Table \ref{tab:Int_ranges}).
Top panel: the mean spectrum, as
defined in Equation (\ref{eq:meanspectrum}), is shown as a black solid line. The spectral model described in
Section \ref{sec:model} is shown in red. The deep trough centered at 1215\,\AA\ is Galactic \Lya\ absorption
and the strong narrow emission line at the center of this trough is geocoronal \Lya\ emission.
The narrow Galactic absorption lines, although generally saturated, are never black at line center because the
thermal broadening of this cold gas is still far below the resolution of COS.
Bottom panel: the RMS spectrum, as defined in Equation (\ref{eq:rmsspectrum}), is shown as a black solid line, 
while the intrinsic RMS spectrum $\sigma_0$ (see Section \ref{sec:MeanRMS}) is shown in blue. 
In grey we show the total error on the mean, which is the statistical error  (Equation \ref{eq:staterror}) 
combined in quadrature with the fractional error in precision, 
$\delta_{\rm P} = 1.1$\% for the G130M spectra.
Note the difference in the flux scale between the two panels.}
\label{fig:Mean_rms_G130}
\end{figure*}

\begin{figure*}
\centering
\includegraphics[width=0.9\textwidth]{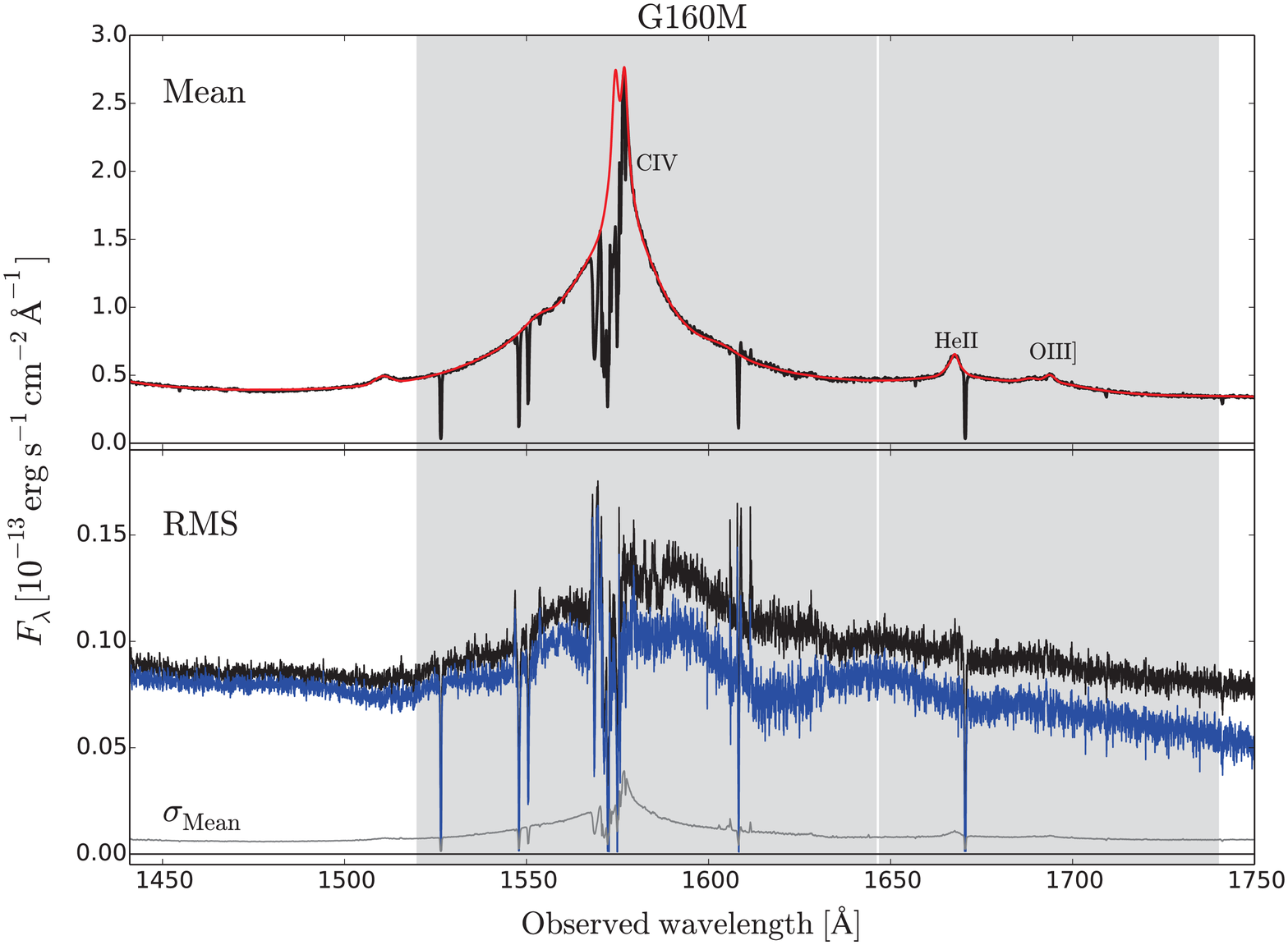}
\caption{Spectra obtained with the G160M grating. The shaded areas show the integration regions
defined in Table \ref{tab:Int_ranges}). Top panel: the mean spectrum, 
as defined in Equation (\ref{eq:meanspectrum}), is shown as a black solid line. The spectral model described in
Section \ref{sec:model} is shown in red. 
The narrow Galactic absorption lines, although generally saturated, are never black at line center because the
thermal broadening of this cold gas is still far below the resolution of COS.
Bottom panel: the RMS spectrum, as defined in Equation (\ref{eq:rmsspectrum}), is shown as a black solid line, 
while the intrinsic RMS spectrum $\sigma_0$ (see Section \ref{sec:MeanRMS}) is shown in blue. 
In grey we show the total error on the mean, which is the statistical error  (Equation \ref{eq:staterror}) 
combined in quadrature with the fractional error in precision, 
$\delta_{\rm P} = 1.4$\% for the G160M spectra.
Note the difference in the flux scale between the two panels.}
\label{fig:Mean_rms_G160}
\end{figure*}

The total uncertainty in the mean spectra consists of this statistical
uncertainty and our estimate of the fractional uncertainty in
precision as described above, which amounts to 
$\delta_{\rm P}({\rm G130M}) \approx 1.1\%$ and 
$\delta_{\rm P}({\rm G160M}) \approx
1.4\%$.  To determine the total uncertainty, the statistical
uncertainty (Equation \ref{eq:staterror}) and the uncertainty in
precision ($\delta_{\rm P} \overline{F}(\lambda)$) are added in
quadrature.  
The mean and RMS spectra for the G130M and G160M
settings are shown in Figures \ref{fig:Mean_rms_G130} and
\ref{fig:Mean_rms_G160}, respectively.

The RMS spectrum resulting from Equation (\ref{eq:rmsspectrum})
combines both the intrinsic variability and the variance
due to noise, as discussed by \cite{Park12} and \cite{Barth15}.
In order to isolate the RMS spectrum of the intrinsic 
variations $\sigma_0$, we
model the distribution of the residuals of each pixel about
the mean. The combined statistical and systematic noise in each
pixel of spectrum $i$ is thus 
$(\sigma_i^2+(\delta_{\rm P} F_i)^2)^{1/2}$. Assuming that the
flux measurement errors and the intrinsic variations arise from
independent Gaussian random processes, 
we find maximum likelihood estimates for 
the optimal average\footnote{ The unweighted average $\overline{F}$ in
Equation \ref{eq:meanspectrum} is formally distinct from
the optimal average $\overline{F_\mu}$, though practically they are
indistinguishable for these data.},
$\overline{F_\mu}$, and $\sigma_0$
by minimizing
\begin{equation}
        -2\,\ln{ L \left( \overline{F_\mu}, \sigma_0 \right) }
        = \chi^2 + \sum_{i=1}^N
        \ln{ \left[ \sigma^2_0 + \sigma^2_i + 
\left( \delta_{\rm P} F_i \right)^2 \right] }
\ ,
\end{equation}
where
\begin{equation}
        \chi^2 = \sum_{i=1}^N
        \frac{ \displaystyle \left( F_i - \overline{F_\mu} \right)^2 }
        { \displaystyle \sigma^2_0 + \sigma^2_i + 
\left( \delta_{\rm P} F_i \right)^2 }
\ ,
\end{equation}
and $\delta_{\rm P}  F_i$ is the precision of the $i$th spectrum.
This estimate of the intrinsic RMS spectrum ($\sigma_0$) is also shown in 
Figures \ref{fig:Mean_rms_G130} and \ref{fig:Mean_rms_G160}.

\subsection{Integrated Light Curves}
\label{sec:intlc}

The next step in our initial analysis is to produce light curves for the continuum and
emission lines. At this stage, our goal is to make simple measurements from the reduced
spectra, introducing as few assumptions as possible. 
All flux measurements are performed on spectra in the observed frame. We have not
corrected the spectra for Galactic extinction in order to facilitate the cleanest comparison
with other measurements to be reported elsewhere in this series of papers 
(e.g., broad-band photometry).

There are bad pixels throughout the spectrum, and their location and severity
change with time, instrument settings, and airglow
subtraction (e.g., if a spectrum is taken entirely in orbital bright
time, the flux in the airglow windows is set to zero and the pixels are
flagged as bad pixels). To prevent the introduction of artificial
variations in the relative flux estimates, bad pixels are masked 
throughout the dataset. This means that if a pixel is bad in any
of the visits,  the pixel is
masked out in each of the 171 spectra.  We further mask 
Galactic \Lya\ absorption and airglow region.  Integration ranges (listed in
Table \ref{tab:Int_ranges}) were chosen using the mean spectra
in Figures \ref{fig:Mean_rms_G130} and \ref{fig:Mean_rms_G160}
as a guide.  Continuum ranges are chosen to
be as uncontaminated as possible by absorption lines and
broad emission-line wings.  In the case of overlapping emission lines
(e.g., \civ \ and \heii), the boundary wavelength corresponds to the
wavelength at which the fluxes of the two lines are comparable.
We do not mask absorption lines at
this stage in our analysis. We are unable to cleanly separate
\nv\ and \Lya\ using this simple procedure.

\begin{deluxetable*}{lccc}
\tablecaption{Integration Limits for Light Curves}
\tablehead{\colhead{Emission} & \colhead{Integration} & 
\colhead{Shortward} & \colhead{Longward}\\
\colhead{Component} & \colhead{Limits} & \colhead{Continuum Region} 
& \colhead{Continuum Region}}
\startdata
$F_{\lambda} \left( {\rm 1367 \AA} \right) $ & 1364.5--1369.5 & -- & -- \\
\Lya\,$\lambda 1215$ & 1201.0--1255.0 & 1155.0--1160.0 & 1364.5--1369.5 \\
\siiv\,$\lambda1400$\tablenotemark{a} & 1405.0--1455.0 & 1364.5--1369.5 & 1460.0--1463.5 \\
\civ\,$\lambda1549$ & 1520.0--1646.0 & 1475.0--1482.0 & 1743.0--1749.0 \\
\heii\,$\lambda1640$\tablenotemark{b} & 1647.0--1740.0 & 1475.0--1482.0 & 1743.0--1749.0
\enddata
\tablecomments{\label{tab:Int_ranges} All regions are in the observed frame (\AA).}
\tablenotetext{a}{Integration range also includes O\,{\sc iv}]\,$\lambda 1402$.}
\tablenotetext{b}{Integration range also includes O\,{\sc iii}]\,$\lambda 1663$.}
\end{deluxetable*}

Continuum fluxes are measured as the weighted mean of the flux density in the integration region, with weights equal to the inverse of the variance,  
\begin{equation}
F_\lambda = \left( \sum_{i=1}^N{w_i \ F_i}\right) \left( \sum_{i=1}^N{w_i}\right)^{-1},
\end{equation}
where $w_i=\sigma_{F_i}^{-2}$ as in Equation (\ref{eq:staterror}).
Statistical uncertainties computed by {\tt CalCOS} are corrected for low counts following \citet{Gehrels1986}. 
Statistical uncertainties on the mean fluxes are obtained through standard error propagation, 
\begin{equation}
\label{eq:meanstaterror}
 \sigma_{F_\lambda} = \left( \sum_{i=1}^N{w_i} \right)^{-1/2}.
 \end{equation}
In all cases, bad pixels are excluded from the computation.

Emission-line fluxes are measured as the numerical integral of the
emission flux above a locally defined continuum defined by the
relatively featureless windows given in Table \ref{tab:Int_ranges}.
To estimate the local continuum underneath the line we performed a
$\chi^2$ linear fit of the continuum flux in the selected regions. The
linear local continuum is then subtracted from the emission component,
again masking bad pixels. The line flux is numerically
integrated over the integration limits given in
Table \ref{tab:Int_ranges} using Simpson's method. We do
not interpolate over bad pixels. We note, however, that the  difference
between integrating over the bad pixels and computing the integral
excluding them is $< 0.1$\%. Statistical errors are computed numerically
by creating $N_{\rm sample}=5000$ realizations of the line flux and the
underlying linear continuum. The
flux $F_\lambda$ is randomly generated from a Gaussian
distribution having mean equal to the flux of the spectral element and
width $\sigma$ equal to the statistical error on the flux.
For the linear continuum,
we generate $N_{\rm sample}$ fits having a mean equal to
the best fit values and covariance equal to their covariance
matrix. For each realization, a line-flux estimate is then obtained by
subtracting the linear continuum and by performing the numerical
integration of the residuals. Confidence levels ($1 \sigma$) are
finally obtained from the distribution of the $N_{\rm sample}$ line
fluxes. When the error bars are asymmetric, we adopt the larger error
as the statistical error associated with the integrated flux.

As noted above, we adopt as the fractional error in precision 
$\delta_{\rm P} = 1.1\%$  and $\delta_{\rm P} = 1.4\%$ for the G130M and G160M settings, respectively.
This is added in quadrature to statistical error of the integrated fluxes.
The error in the precision dominates throughout the G160M spectra and in the
G130M spectra as well, except at wavelengths shortward of $\sim1180$\,\AA\, longward of
$\sim1425$\,\AA, and in the core of the \Lya\ complex.

\begin{figure*}
\centering
\includegraphics[width=0.9\textwidth]{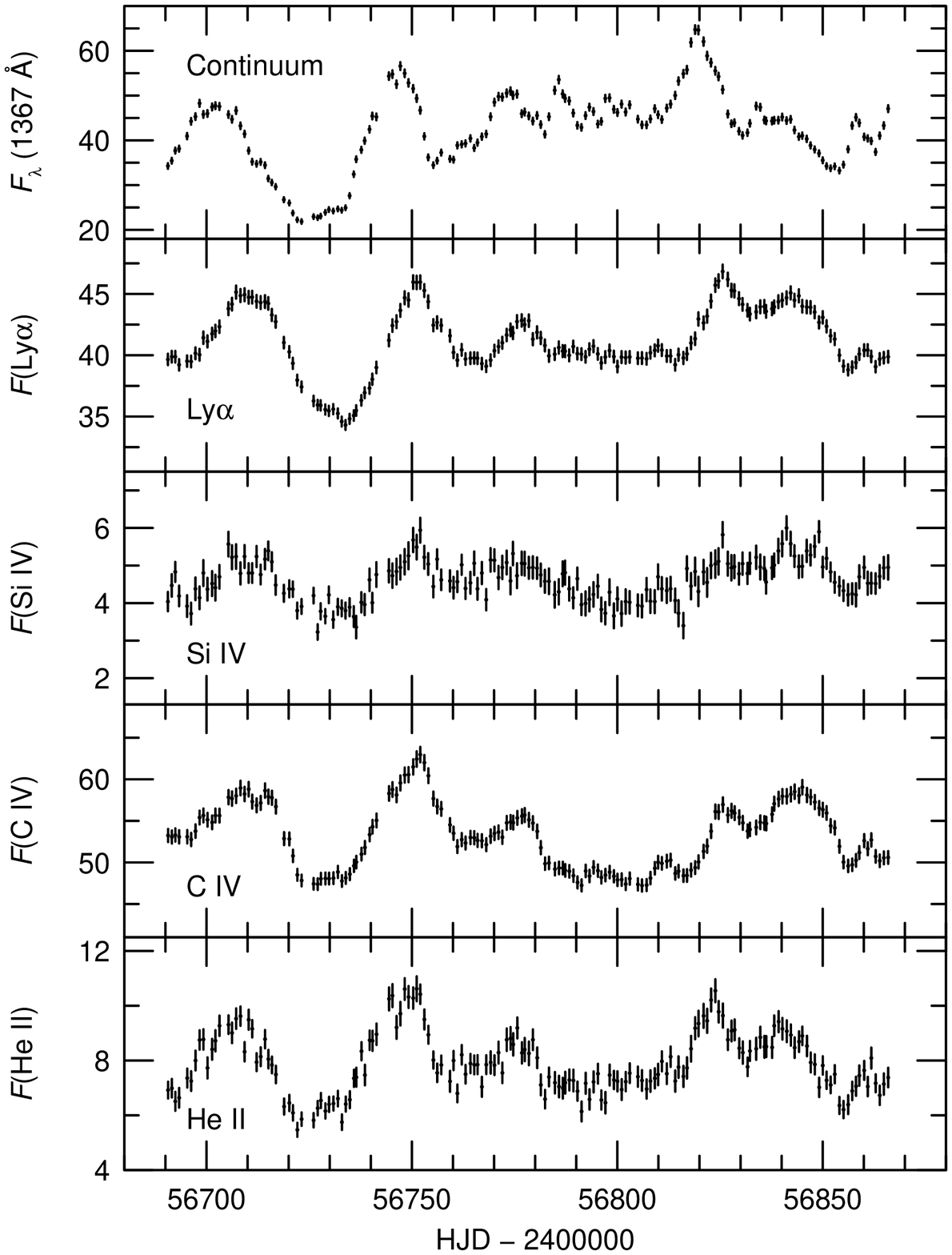}
\caption{Integrated light curves. The continuum flux at 1367\,\AA\
is in units of $10^{-15}$ erg s$^{-1}$ cm$^{-2}$ \AA$^{-1}$ and
the line fluxes are in units of $10^{-13}$ erg s$^{-1}$ cm$^{-2}$ and are in the observed frame. 
Flux uncertainties include both statistical and systematic errors.}
\label{fig:Light_curves}
\end{figure*}

The final continuum and emission-line light curves are listed in
Table \ref{tab:light_curves} and shown in Figure
\ref{fig:Light_curves}. The light curve statistics are given in Table
\ref{tab:lc_stats}.  The average interval between two consecutive
observations is $\langle \Delta t \rangle = 1.0$ days with an rms
$\sigma (\Delta t) = 0.3$ days. The median interval between observations
is $\Delta t_{\rm med} =1.0$ days. The largest gaps between consecutive
observations are three days (on two occasions) and two days (on six occasions).

\begin{deluxetable*}{cccc|ccc}
\tablecaption{Continuum and Emission-Line Light Curves}
\tablehead{\multicolumn{4}{c}{G130M} & \multicolumn{3}{c}{G160M} \\
\hline
\colhead{HJD\tablenotemark{a}} & \colhead{$F_\lambda\left({\rm 1367 \AA}\right)$\tablenotemark{b}}& \colhead{$F$(\Lya)\tablenotemark{c}}  & 
 \colhead{$F$(\siiv)\tablenotemark{c}} & \colhead{HJD\tablenotemark{a}} & \colhead{$F$(\civ)\tablenotemark{c}} & \colhead{$F$(\heii)\tablenotemark{c}}}
\startdata
6690.6120 & $34.27\pm0.64$ & $39.66\pm0.47$ & $ 4.04 \pm 0.26$ & $6690.6479$ &$ 53.24 \pm 0.79$ &$ 6.92 \pm 0.31$ \\
6691.5416 & $35.45\pm0.65$ & $39.88\pm0.48$ & $ 4.47 \pm 0.30$ & $6691.5760$ &$ 53.06 \pm 0.79$ &$ 6.99 \pm 0.34$ \\
6692.3940 & $37.71\pm0.67$ & $39.88\pm0.48$ & $ 4.83 \pm 0.27$ & $6692.4084$ &$ 53.30 \pm 0.80$ &$ 6.51 \pm 0.35$ \\
6693.3237 & $38.14\pm0.68$ & $39.22\pm0.47$ & $ 4.19 \pm 0.28$ & $6693.3380$ &$ 53.08 \pm 0.80$ &$ 6.64 \pm 0.36$ \\
6695.2701 & $40.94\pm0.71$ & $39.52\pm0.47$ & $ 3.92 \pm 0.29$ & $6695.3145$ &$ 53.09 \pm 0.81$ &$ 7.36 \pm 0.35$ \\
6696.2459 & $44.25\pm0.75$ & $39.49\pm0.48$ & $ 3.72 \pm 0.29$ & $6696.2602$ &$ 52.76 \pm 0.80$ &$ 7.25 \pm 0.36$ \\
6697.3080 & $45.30\pm0.75$ & $40.16\pm0.49$ & $ 4.38 \pm 0.30$ & $6697.3223$ &$ 53.77 \pm 0.82$ &$ 8.00 \pm 0.36$ \\
6698.3041 & $48.27\pm0.79$ & $40.04\pm0.48$ & $ 4.14 \pm 0.30$ & $6698.3184$ &$ 55.40 \pm 0.83$ &$ 8.75 \pm 0.36$ \\
6699.2338 & $45.80\pm0.76$ & $41.43\pm0.51$ & $ 4.80 \pm 0.35$ & $6699.2481$ &$ 55.65 \pm 0.84$ &$ 8.77 \pm 0.37$ \\
6700.2299 & $46.00\pm0.76$ & $41.13\pm0.50$ & $ 4.37 \pm 0.30$ & $6700.2442$ &$ 55.13 \pm 0.83$ &$ 7.73 \pm 0.38$ \\
6701.3588 & $47.46\pm0.78$ & $41.75\pm0.50$ & $ 4.52 \pm 0.33$ & $6701.3731$ &$ 54.82 \pm 0.83$ &$ 8.41 \pm 0.38$ \\
6702.1557 & $47.74\pm0.78$ & $41.98\pm0.51$ & $ 4.41 \pm 0.34$ & $6702.1700$ &$ 55.64 \pm 0.84$ &$ 8.72 \pm 0.39$ \\
6703.1518 & $47.56\pm0.78$ & $42.33\pm0.51$ & $ 4.70 \pm 0.32$ & $6703.1661$ &$ 55.63 \pm 0.84$ &$ 9.27 \pm 0.37$ \\
6705.3432 & $45.77\pm0.76$ & $43.80\pm0.53$ & $ 5.57 \pm 0.33$ & $6705.3575$ &$ 57.85 \pm 0.86$ &$ 9.31 \pm 0.34$
\enddata
\tablecomments{\label{tab:light_curves} Full table is given in the published version. Integrated light curves in the observed frame.
Flux uncertainties include both statistical and systematic errors.}
\tablenotetext{a}{Midpoint of the observation $({\rm HJD} - 2450000).$}
\tablenotetext{b}{Units of 10$^{-15}$ erg s$^{-1}$ cm$^{-2}$ \AA$^{-1}$.}
\tablenotetext{c}{Units of 10$^{-13}$ erg s$^{-1}$ cm$^{-2}$. }
\end{deluxetable*}

\begin{deluxetable*}{lcccccc}
\tablecaption{Light Curve Statistics}
\tablehead{\colhead{Emission} & \colhead{Mean and} & \colhead{Mean} & \colhead{ } & 
\colhead{Maximum} & \colhead{Minimum} & \colhead{ }\\
\colhead{Component} & \colhead{RMS Flux} & \colhead{Fractional Error} & \colhead{$F_{\rm var}$\tablenotemark{a}} & \colhead{Flux} & \colhead{Flux} & \colhead{$R_{\rm max}$\tablenotemark{b}}\\
\colhead{(1)} & \colhead{(2)} & \colhead{(3)} & \colhead{(4)} & \colhead{(5)} & \colhead{(6)} & \colhead{(7)}}
\startdata
$F_\lambda\left({\rm 1367 \AA}\right)$\tablenotemark{c} & $42.64 \pm 8.60$ & 0.017 & 0.201 & 64.74 & 21.87 & $2.96 \pm 0.08$ \\ 
$F$(\Lya)\tablenotemark{d} & $41.22 \pm 2.71$ & 0.012 & 0.065 & 46.84 & 34.31 & $1.37 \pm 0.02$ \\
$F$(\siiv)\tablenotemark{d} & $\phantom{4}4.62 \pm 0.55$ & 0.065 & 0.099 & 6.00 & 3.23 & $1.86 \pm 0.15$ \\ 
$F$(\civ)\tablenotemark{d} & $53.28 \pm 3.91$ & 0.015 & 0.072 & 62.97 & 47.23 & $1.33 \pm 0.03$ \\
$F$(\heii)\tablenotemark{d} & $\phantom{4}7.93 \pm 1.13$ & 0.046 & 0.135 & 10.62 & 5.47 & $1.94 \pm 0.12$
\enddata
\tablecomments{\label{tab:lc_stats}Light curves statistics are in the observed frame.}
\tablenotetext{a}{Excess variance, defined as 
\begin{equation}
F_{\rm var} = \frac{\sqrt{\sigma^2-\delta^2}}{\langle F \rangle} \,
\end{equation} 
where $\sigma$ is the RMS of the observed fluxes (column 2), $\delta$ is the mean 
statistical uncertainty (column 3 times $\langle F \rangle$), and  $\langle F \rangle$ is the mean flux 
in column (2) \citep{Rodriguez97}.}
\tablenotetext{b}{Ratio between maximum and minimum flux.}
\tablenotetext{c}{Units of 10$^{-15}$ erg s$^{-1}$ cm$^{-2}$ \AA$^{-1}$.}
\tablenotetext{d}{Units of 10$^{-13}$ erg s$^{-1}$ cm$^{-2}$.}
\end{deluxetable*}

\subsection{Time-Series Measurements}
\label{sec:time_series}

Certain simplifying assumptions underlie the RM technique. Most
time-series analyses start with the assumption that the emission-line
light curves are simply scaled, time-delayed, and possibly smoothed
versions of the continuum light curve. Inspection of the light curves
in Figure \ref{fig:Light_curves} suggests that this is an entirely
reasonable assumption for the first half of the campaign. However,
approximately halfway through the campaign, the emission-line response
becomes more complicated.  Between approximately HJD2456780 and
2456815, the emission-line light curves are either flat (\Lya, \heii)
or decreasing (\civ) while the continuum is slowly rising. Moreover,
the intensity ratio between the last two strong peaks in the continuum
light curve at around HJD2456820 and 2456840 seems to be almost
inverted in the lines, with the second peak being stronger than
the first one (especially in \civ). There is also a small event in the
continuum light curve around HJD2456785 that does not appear to have
counterparts in the emission-line light curves.
The line light curve that seems to best trace
the continuum is the \heii \ light curve, which is sensitive
to the continuum at energies above 4 Ryd. It is also the only strong
line in the COS spectra that is neither a resonance line nor
self-absorbed. Moreover, it is the line that arises closest to the continuum
source, as we will show below.

Because of the changing character of the emission-line response, 
for our initial analysis we measure emission-line lags (a) for the entire
data set and (b) for subsets that divide the data into two separate halves of 85 observations each. The first subset,
which we will refer to as ``T1,'' runs from HJD2456690 to 2456780 and the second subset, ``T2,'' runs from
HJD2456781 to 2456865.

We first measured the emission-line lags relative to the continuum
variations by cross-correlation of the light curves.  We used the
interpolation cross-correlation (ICCF) method as implemented by
\cite{Peterson04}. In this method, uncertainties are estimated using a
model-independent Monte Carlo method referred to as ``flux
randomization and random subset selection (FR/RSS).'' For each
realization, $N$ data points are selected from a light curve with $N$
independent values, without regard to whether or not any particular
point has been previously selected. For data points selected $n$
times in a given realization, the flux error associated with that data
point is reduced by a factor of $n^{1/2}$. The flux measured at each
data point is then altered by adding or subtracting a random Gaussian
deviate scaled by the flux uncertainty ascribed to that point.  Each
realization yields a cross-correlation function that has a maximum
linear correlation coefficient $r_{\rm max}$ that occurs at a lag
$\tau_{\rm peak}$. We also compute the centroid $\tau_{\rm cent}$ of
the cross-correlation function using all the points near $\tau_{\rm
  peak}$ with $r(\tau) \geq 0.8\,r_{\rm max}$.  Typically a few thousand
realizations are used to construct distribution functions for the
ICCF centroid and peak. We adopt the median values of the cross-correlation
centroid distribution and the cross-correlation peak distribution as
our lag measurements. The uncertainties, which are not necessarily
symmetric, correspond to a 68\% confidence level. In general, $\tau_{\rm cent}$ is
found to be a more reliable indicator of the BLR size than $\tau_{\rm peak}$, 
though we record both. The ICCF measurements
of $\tau_{\rm peak}$ and $\tau_{\rm cent}$ for the four strongest UV
emission lines are given in the second and third columns,
respectively, in Table \ref{tab:lags}.

\begin{deluxetable*}{lccccc}
\tablecaption{Emission-Line Lags}
\tablehead{\colhead{Emission Line} & \colhead{$\tau_{\rm peak}$\tablenotemark{a}} & \colhead{$\tau_{\rm cent}$\tablenotemark{a}} & \colhead{$\tau_{\tt JAVELIN}$\tablenotemark{a}} & \colhead{$\tau_{\rm cent, T1}$\tablenotemark{b}} & \colhead{$\tau_{\rm cent, T2}$\tablenotemark{c}}}
\startdata
\Lya & $6.1^{+0.4}_{-0.5}$&$6.19^{+0.29}_{-0.25}$&$5.80_{-0.39}^{+0.36}$&$5.90^{+0.30}_{-0.29}$&$7.73^{+0.76}_{-0.57}$ \\
\siiv & $5.5^{+1.1}_{-1.1}$&$5.44^{+0.70}_{-0.71}$&$5.94_{-0.55 }^{+0.53}  $&$4.99^{+0.75}_{-0.68}$&$7.22^{+1.33}_{-1.06}$ \\
\civ &  $5.2^{+0.7}_{-0.6}$&$5.33^{+0.44}_{-0.48}$&$4.59_{-0.42}^{+0.68}$&$4.61^{+0.36}_{-0.35}$&$9.24^{+1.04}_{-1.04}$ \\
\heii & $2.4^{+0.3}_{-0.8}$&$2.50^{+0.34}_{-0.31}$&$2.42_{-0.06}^{+0.67}$&$2.11^{+0.43}_{-0.38}$&$3.87^{+0.71}_{-0.58}$
\enddata
\tablecomments{\label{tab:lags} Delays measured in light days in the rest frame of NGC 5548.}
\tablenotetext{a}{Complete dataset: 171 visits}
\tablenotetext{b}{T1 dataset: visits 1--85}
\tablenotetext{c}{T2 dataset: visits 86--171}
\end{deluxetable*}

We have also estimated emission-line lags using {\tt JAVELIN}, 
which is an improved version of {\tt SPEAR}
\citep{Zu11}. {\tt JAVELIN} assumes that the emission-line light
curves are shifted and smoothed versions of the continuum light
curve (as with the ICCF analysis), where the continuum is modeled as a
damped random walk \citep{Kelly09,Kozlowski10,MacLeod10}
with uncertainties determined using the
Markov Chain Monte Carlo method. We model the full dataset, and each
line light curve was run independently with the continuum.  The
results, given in column (4) of Table \ref{tab:lags}, are in good
agreement with the ICCF analysis, as expected.

\begin{figure*}
\centering
\includegraphics[width=0.9\textwidth]{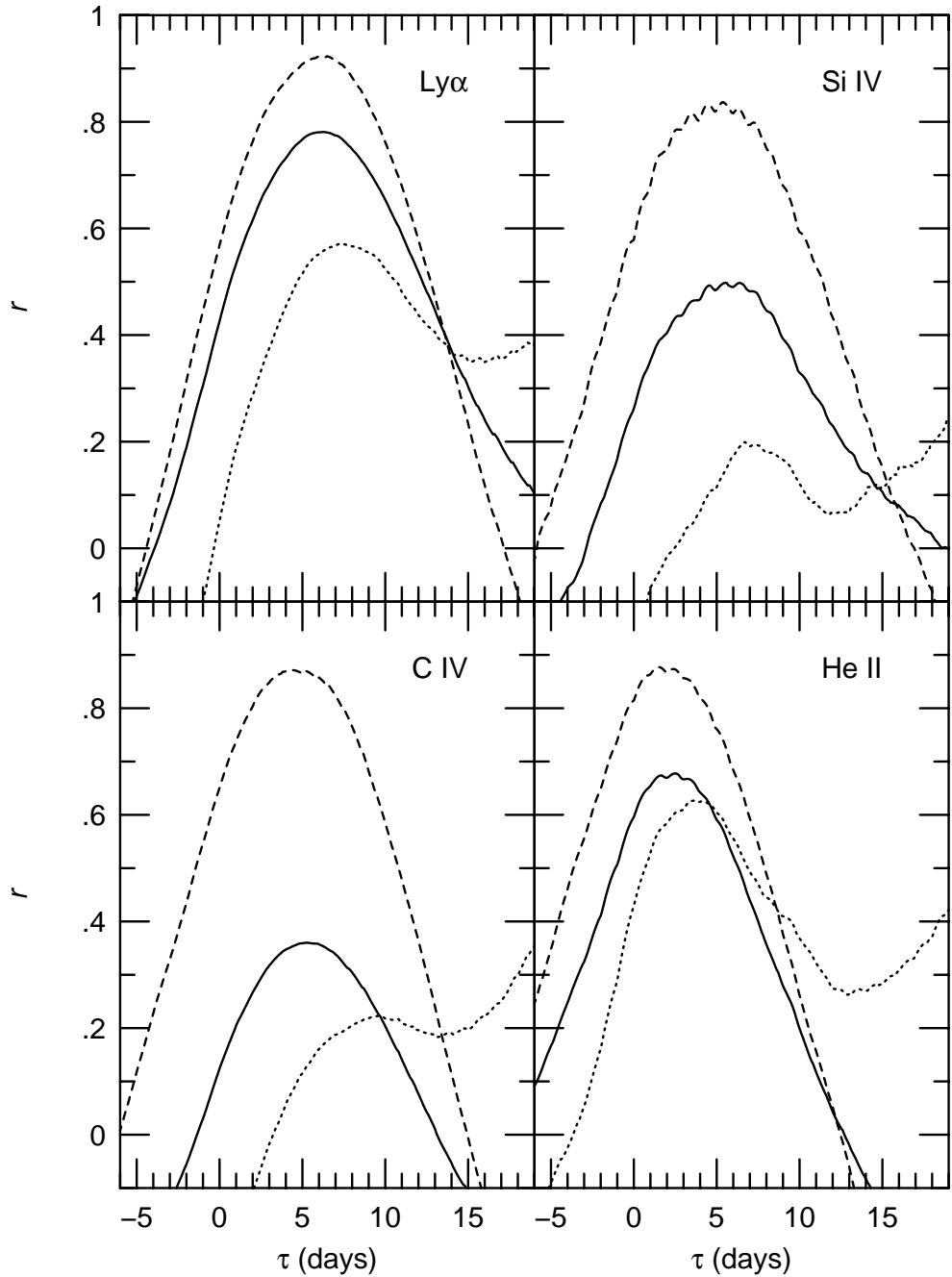}
\caption{Interpolated cross-correlation functions (ICCFs) computed by cross-correlating each emission-line
light curve with the 1367\,\AA\ continuum light curve. The solid line is the ICCF for the entire data set.
The dashed line is for the T1 subsample (the first 85 visits) and the dotted line is for the T2 subsample
(the last 85 visits).}
\label{fig:CCF}
\end{figure*}

In columns (5) and (6) of Table \ref{tab:lags}, we also give the ICCF centroid values for the T1 and T2 subsets.
We also show the ICCFs for the entire sample and the T1 and T2 subsamples in Figure \ref{fig:CCF}.
In general, the lags for the T1 subsample have the smallest uncertainties and the ICCFs have the largest
peak correlation coefficients $r_{\rm max}$, as expected. The T2 subset, on the other hand, yields 
lags with larger uncertainties and ICCFs with lower values of $r_{\rm max}$ (indeed, {\em much} lower
in the case of \siiv\ and \civ), again as expected from visual inspection of the light curves. The T2 lags
are also larger than those from T1, probably only in small part because the continuum is on average brighter 
(by $\sim15$\% on average) during the
second half of the campaign so the BLR gas that is most responsive to continuum changes is farther 
away from the central source.

\subsection{Velocity-Binned Results}
As noted earlier (Section \ref{sec:design}), this program was designed to recover kinematic information
about the BLR by resolving the emission-line response as a function of radial velocity. This will be
explored in detail in subsequent papers in this series. 
Here we carry out a simple preliminary analysis intended to show only whether 
velocity-dependent information is present in the data. We isolate the \Lya\ and \civ\ profiles as
described earlier (Section \ref{sec:intlc}) but then integrate the fluxes in bins of width
500\,\kms, except in the shortward wing of \Lya\  ($-10,000\,\kms \leq \Delta V \leq -7000\,\kms$) where we use
1000\,\kms\ bins on account of the low flux in the blue wing of this line.

\begin{figure*}
\centering
\includegraphics[width=0.9\textwidth]{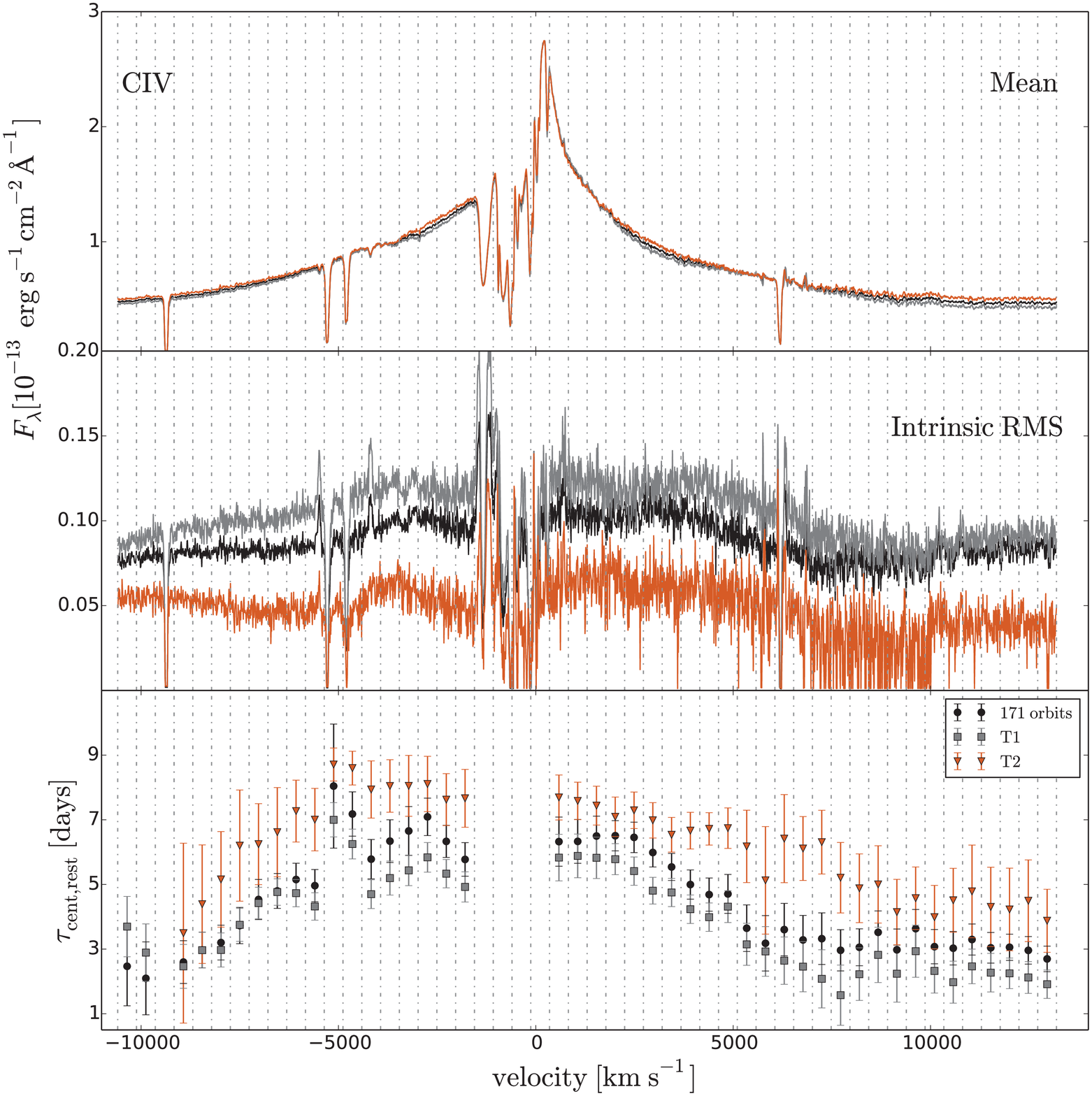}
\caption{Velocity-binned results for the \civ\,$\lambda1549$ emission line. 
The top and middle panels show the mean and
intrinsic RMS ($\sigma_0$) spectra, respectively. The bottom panel shows the centroid of the cross-correlation function for each individual velocity bin,
corrected to the rest frame of NGC\,5548. Velocity bins with no lag measurement contain too little total flux
to reliably characterize the variations.
In each case, the black line represents the entire dataset, and the T1 (first 85 visits) and T2 (last 85 visits) subsets 
are shown in gray and orange, respectively. Note that the shortest lags are found for the highest-velocity gas.}
\label{fig:CIV_bin_T1T2}
\end{figure*}

We show the ICCF centroids for each velocity bin in the \civ\ emission-line profile in 
the bottom panel of Figure \ref{fig:CIV_bin_T1T2}.  For the two subsets as well as the entire dataset, 
we see that there is a clear ordered structure in the kinematics. We cannot infer much from such a simple analysis,
of course, because we cannot accurately characterize a complex velocity field with a single number.
It is reassuring, however, that the general pattern is similar to what has been seen in other
objects and is qualitatively consistent with a virialized region (i.e., the high velocity wings respond first).
The middle panel of Figure \ref{fig:CIV_bin_T1T2} shows the RMS spectra for the entire dataset 
and the two subsets. 

\begin{figure*}
\centering
\includegraphics[width=0.9\textwidth]{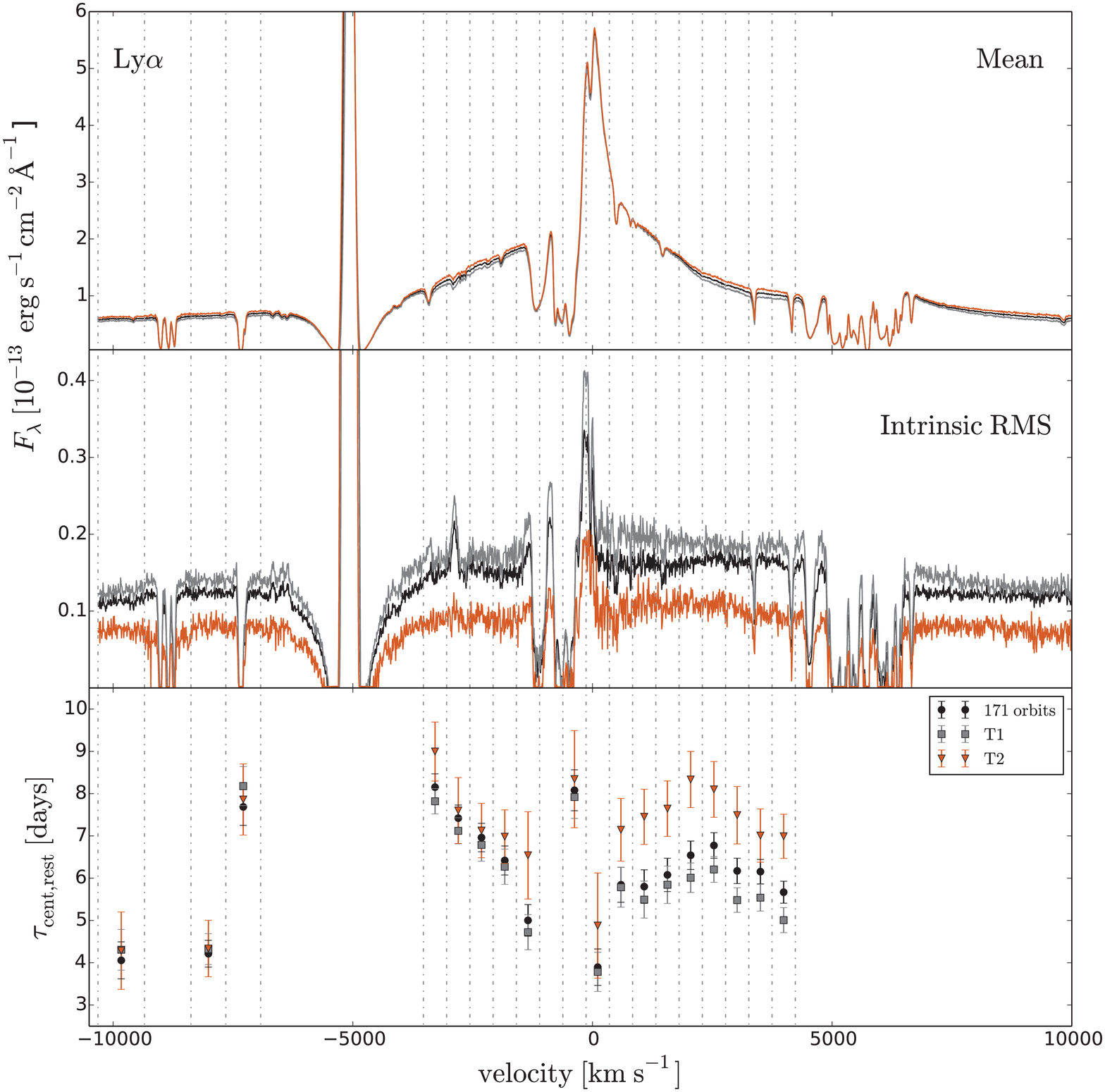}
\caption{Velocity-binned results for the \Lya\,$\lambda1215$ emission line. 
The top and middle panels show the mean and
intrinsic RMS ($\sigma_0$) spectra, respectively. The bottom panel shows the centroid of the cross-correlation function for each individual velocity bin,
corrected to the rest frame of NGC\,5548. Velocity bins with no lag measurement contain too little total flux
to reliably characterize the variations.
In each case, the black line represents the entire dataset, and the T1 (first 85 visits) and T2 (last 85 visits) subsets 
are shown in gray and orange, respectively. The large gap at around $-5000\,\kms$ avoids the region of the spectrum affected
by geocoronal emission and Galactic absorption. }
\label{fig:Lya_bin_T1T2}
\end{figure*}

In the bottom panel of Figure \ref{fig:Lya_bin_T1T2}, we show the velocity-binned ICCF centroids
for the \Lya\ emission line. Again, a clear pattern emerges, as the lags in each velocity bin are highly
correlated with those of adjacent bins. However, the pattern that emerges is unlike what is seen in \civ; the 
largest lags are at intermediate velocities and the lags decrease toward line center. However, given the severe
blending and strong absorption, detailed modeling will be required before any meaningful conclusions can be drawn.

\subsection{Modeling the mean spectrum}
\label{sec:model}
The mean spectra shown in Figures \ref{fig:Mean_rms_G130} and \ref{fig:Mean_rms_G160}
can be compared with earlier UV spectra of NGC 5548 obtained with \HST\
\citep[e.g.,][]{Korista95,Kaastra14}. The earliest high-quality UV spectra of NGC 5548 showed
only weak absorption in the resonance lines
(\Lya, \nv, \siiv, and \civ),
a factor that contributed to our selection
of NGC 5548 as a target for this investigation. The 2013 spectra \citep{Kaastra14}, however,
revealed not only several strong narrow absorption features in the resonance lines,
but also evidence for a relatively large ``obscurer'' that strongly absorbs the emission in the blue
wings of the resonance lines. This feature is still present in our spectra obtained a year 
later, but is weaker than it was in 2013.
¯¯
The presence of this absorption, combined with the blending of various emission
features (\Lya\ and \nv, \siiv\ and \oiv], \civ\ and \heii),
complicates analysis of these spectra.
To characterize the emission-line structure of NGC 5548 and  
guide our selection of continuum windows for our emission-line flux
measurements, we fit a heuristic model to the lines and continuum in the
mean spectra.
Our model for NGC 5548 is similar to that adopted by \cite{Kaastra14},
and it includes the broad absorption features associated with all permitted
transitions in the spectrum.
Our adopted continuum is a reddened power law of the form
$F_\lambda(\lambda)  = F_\lambda(1000\, {\rm \AA}) (\lambda / 1000\,{\rm \AA})^{- \alpha}$.
We correct for $E(\bv) = 0.017$\,mag of Galactic extinction \citep{Schlegel98,Schlafly11}
using the prescription of \cite{Cardelli89} and 
$R_V = 3.1$.
We do not apply any correction for possible internal extinction in NGC\,5548.
Longward of 1550\,\AA, we also include blended \feii \ emission as modeled by
\cite{Wills85}, broadened with a Gaussian with full-width at half-maximum 
${\rm FWHM} = 4000\,{\rm km~s}^{-1}$.
We model the emission lines with multiple Gaussian components.
These are not an orthogonal set, and the decomposition is not rigorously
unique, but they characterize each line profile well.

For the brightest lines, we start with a narrow component, typically with
${\rm FWHM} \approx 300\,{\rm km~s}^{-1}$.
This component is essentially identical to the narrow component  used by
\cite{Crenshaw09} for fitting the 2004 STIS spectrum of NGC 5548.
Since this narrow component is difficult to deblend from the broader
components of each line, and since \cite{Crenshaw09} saw little variation in
narrow-line intensity over time, we fix the flux and widths of the narrow
components of \Lya, \nv, \civ, and \heii \ to the values we used to fit the
2004 STIS spectrum. The \siiv \ lines do not have a detectable narrow-line component.
We note that while \cite{PetersonPlus13} detected changes in the
strength of the narrow [\oiii]\,$\lambda\lambda4959$, 5007 lines
over timescales of years, the variations over the last decade have
been only at the few percent level. Similar variations in the
narrow components of the UV lines would not be easily detected
here because the narrow components are all so weak.

Next we add an intermediate-width component with
${\rm FWHM} \approx 800\,{\rm km~s}^{-1}$ without ascribing physical
meaning to it,  using the STIS 2004 spectrum as a model. 
An intermediate-width component is
included in \Lya, \nv, \siiv, \civ, and \heii, as well as in the
fainter lines \ciiiex\,$\lambda$1176, \Sizw\,$\lambda$1260,
\Sizw$+$\oi\,$\lambda$1304, \cii\, $\lambda$1335,
\nivsf\,$\lambda$1486, \oiiisf\,$\lambda$1663, and \niiisf\,$\lambda$1750.
For the weaker lines, this is often the only component detected, so its
flux, width, and position are all allowed to vary.
For the stronger lines, as for the NLR components, we again keep the fluxes and widths
of the intermediate-width components fixed at the values found for the STIS 2004
spectrum since \cite{Crenshaw09} found that these components vary only
slightly in flux over several years.
For the stronger lines, we next include broader components with
$ {\rm FWHM} \approx  3000,$ 8000, and $15,000\,{\rm km~s}^{-1}$, respectively.
For the doublets of \nv, \siiv, and \civ, we assume the line-emitting gas is optically
thick and fix
the flux ratio of each pair to 1:1, 
although for the $15,000\,{\rm km~s}^{-1}$ component,
only a single Gaussian is used.
Finally, as can be seen in the RMS spectrum  in Figure \ref{fig:Mean_rms_G160}, 
there are two weak
bumps that appear on the red and blue wings of the \civ\ emission-line profile
at $\sim1554$\,\AA\ and $\sim1604$\,\AA\ in the observed frame.
These bumps are also present in the mean spectrum; we include a single
Gaussian component to account for each of these bumps.

As described by \cite{Kaastra14}, we use an asymmetric Gaussian with negative flux to model
the broad absorption troughs. The asymmetry in these Gaussian profiles
is introduced by specifying a larger dispersion on the blue side of line
center than on the red side. The asymmetry is fitted as a free parameter,
and the resultant absorption line has a roughly rounded triangular shape
with a blue wing extending from the deepest point in the absorption profile
\citep[for an illustration, see] []{Kaastra14}.
During the first part of our reverberation campaign, when these absorption
features were strongest, an additional depression appeared on the high-velocity
blue tail of the main absorption trough. We use a single, symmetric Gaussian to
model the shape of this additional shallow depression.
For absorption by \nv, \siiv, and \civ, since the individual doublet profiles
are unresolved, we assume the lines are optically thick, so each line in the
doublet has the same strength and profile.

As a final component, we include absorption by
damped Galactic \Lya \ with a column density of
$N({\hi}) = 1.45 \times 10^{20}\,{\rm cm}^{-2}$ \citep{Wakker11}.
The full spectral model for NGC 5548, excluding the narrow absorption, is shown in 
the upper panels of Figure 
\ref{fig:Mean_rms_G130} and Figure \ref{fig:Mean_rms_G160}, superposed on the 
observed mean spectra.\footnote{The complete model of \civ\ showing each of the 
individual components appears in the Supplementary Materials that
accompany \cite{Kaastra14} at
http://www.sciencemag.org/cgi/content/full/science.1253787/DC1. } In future papers, we
will apply this model to individual spectra to isolate the individual emission-line fluxes and
to study absorption-line variability.

\section{Discussion}
To put the results reported here in context, we note that NGC\,5548
has been monitored in the UV for RM purposes on two previous
occasions, as noted in Section \ref{section:introRM}. In 1989,
NGC\,5548 was observed once every four days for eight months with
\IUE\ \citep{Clavel91}. In early 1993, it was observed every other day
with \IUE\ for a period of two months, and during the latter part of
that campaign, it was also observed daily for 39 days with the \HST\
Faint Object Spectrograph \citep{Korista95}. The primary goals of
these two experiments were quite different: the 1989 campaign was the
first massive coordinated RM experiment and it was designed to measure
the mean lags for the strong UV lines. The 1993 campaign was a higher
time-resolution experiment that was designed to eliminate ambiguities
from the 1989 campaign. Specifically, its goals were:
\begin{enumerate}
\item To measure the lag of the most rapidly responding
line, \heii\,$\lambda1640$.
\item To determine whether or not there is a lag
between the UV and optical continuum variations.
\item To determine whether the wings and core of
the \civ\ emission line have different lags.
\end{enumerate}
The first of these goals was met, but 
in the case of the other two, the data only hinted
at results that are being confirmed by this project
(Paper II and Figure \ref{fig:CIV_bin_T1T2}). 

In addition to these RM programs, several 
\HST\ COS spectra of NGC\,5548 were obtained in 2013
with the primary goal of studying absorption features in
the UV as support for an intensive X-ray monitoring program
undertaken with {\em XMM-Newton} \citep{Kaastra14}. Our own
results on variable absorption features constitute an extension of that
effort and will be the subject of a future paper.

Again, for broader context, during the AGN STORM
campaign NGC\,5548 was at about the same
mean continuum luminosity as it was during the 1989 campaign
(but with a somewhat lower amplitude of variability),
somewhat brighter than in the 1993 campaign, and 
decidedly brighter than it was in 2013, which was
near the end of a lower-than-normal state that lasted several
years \citep{PetersonPlus13}. The resonance lines showed
much more self-absorption in this campaign than in either
the 1989 or 1993 observations, but less than seen in 2013.
The emission-line lags were somewhat larger during the 1989
campaign and the emission-line fluxes were higher,
at least in part on account of much lower absorption in
1989. The 1993 emission-line lags were similar to those obtained in
this experiment, but again the line fluxes were larger, but
less self-absorbed.

As already noted, the response of the emission lines becomes
complicated during the second half of the present campaign. The 
\heii\,$\lambda1640$ light curve seems to match the 1367\,\AA\
continuum most closely;
this line responds primarily to continuum emission at $\lambda < 228$\,\AA,
implying that the variations in the 1367\,\AA\ continuum provide
a reasonable proxy for the behavior of the hydrogen-ionizing 
continuum ($\lambda < 912$\,\AA). \heii\ arises closer to the central source than the
other emission lines, and it is also the only non-resonance
line. More detailed analysis will be undertaken once the
\heii\,$\lambda4686$ and Balmer-line results become available
from our contemporaneous ground-based monitoring program
\citep{Pei15}.

In addition to determining the geometry and kinematics of the
BLR, we also wish to use these data to improve on previous estimates of 
the mass of the central black hole.
However, the strong absorption in the blue wings of the 
resonance lines, which was very weak if even present in the
1989 and 1993 campaigns, precludes using simple measurements
of the RMS spectra \citep[e.g.,][]{Peterson04} to make a
mass estimate. More detailed modeling that we hope will
lead to a more accurate black hole mass will be undertaken in
future papers.

To summarize briefly, we have presented the first results from a
UV spectroscopic RM study of NGC\,5548
undertaken with \HST\ COS in 2014.
We detect strong variations in the continuum and find clear delayed
response of the strong emission lines, \Lya, \nv, \siiv, \civ, and 
\heii. A preliminary investigation shows that there is indeed a
strong velocity-dependence of the emission-line lags,
at least in the case of \Lya\ and \civ, although
blending and strong resonance absorption will make
interpretation challenging. However, we have also shown that
a heuristic multicomponent model can account for virtually all
the spectral features. In future contributions, we will use this 
model as a starting point
to explore the UV spectral variations in detail. We will 
also undertake a similar
analysis of contemporaneous optical spectra in an effort to
more completely understand the BLR geometry and kinematics.
In the accompanying Paper II, we combine the continuum
light curve presented here with high-cadence observations with
{\em Swift} for a similar reverberation study of the accretion disk
structure in NGC\,5548.

\acknowledgments 
Support for \HST\ program number GO-13330 was provided by NASA through
a grant from the Space Telescope Science Institute,
which is operated by the Association of Universities for Research in
Astronomy, Inc., under NASA contract NAS5-26555.
We are  grateful for the dedication and hard work by
the Institute staff to make this program a success.
GDR, BMP, CJG, MMF, and RWP are grateful for the support of the
National Science Foundation through grant AST-1008882 to The Ohio
State University. 
AJB and LP have been supported by NSF grant AST-1412693. 
MCB gratefully acknowledges support through NSF CAREER grant AST-1253702 to Georgia State University.
KDD is supported by an NSF Fellowship awarded under grant AST-1302093.
JMG gratefully acknowledges support from NASA under award NNH13CH61C.
PBH is supported by NSERC. 
SRON is financially supported by NWO, the Netherlands Organization for Scientific
Research.
BCK is partially supported by the UC Center for Galaxy Evolution.
CSK acknowledges the support of NSF grant AST-1009756.  
PL acknowledges support from Fondecyt grant \#1120328.
AP acknowledges support from a NSF graduate fellowship and a UCSB Dean's
Fellowship. 
JSS acknowledges CNPq, National Council for Scientific and Technological Development (Brazil)
for partial support and The Ohio State University for warm hospitality.
TT has been supported by NSF grant AST-1412315.
TT and BCK acknowledge support from the Packard Foundation in the form of a Packard Research
Fellowship to TT. TT thanks the American Academy in Rome and thee Observatory
of Monteporzio Catone for kind hospitality.
The Dark Cosmology Centre is funded by the Danish National Research Foundation.
MV gratefully acknowledges support from the Danish Council for Independent Research via grant no.\ DFF – 4002-00275.
This research has made use of
the NASA/IPAC Extragalactic Database (NED), which is operated by the
Jet Propulsion Laboratory, California Institute of Technology, under
contract with the National Aeronautics and Space Administration.
The authors acknowledge with great sadness the loss of our
long-time collaborator in the planning phases of  this project, Professor David J.\ Axon, who  
passed away on 2012 April 5.

\end{document}